\documentclass[vecphys]{svmult}
\usepackage{graphicx,bm,amsmath,amssymb}
\usepackage{color,booktabs}

\usepackage{multicol}        
\usepackage[bottom]{footmisc}

\newcommand{\kf}{k_{\text{F}}}
\newcommand{\vf}{v_{\text{F}}}

\newcommand{\qp}{{\bf q'}}

\newcommand{\zf}{$z_{\kf}$}
\newcommand{\spin}{{\bm \sigma}_1 \cdot {\bm \sigma}_2}
\newcommand{\lm}{\Lambda}

\newcommand{\be}{\begin{equation}}
\newcommand{\ee}{\end{equation}}
\newcommand{\bea}{\begin{eqnarray}}
\newcommand{\eea}{\end{eqnarray}}

\newcommand{\vlowk}{V_{{\rm low}\,k}}

\newcommand{\fmi}{\, \text{fm}^{-1}}
\newcommand{\mev}{\, \text{MeV}}

\begin{document}
\title*{Renormalization group and Fermi liquid theory
for many-nucleon systems}

\author{B.~Friman\inst{1}, K.~Hebeler\inst{2,3} and A.~Schwenk\inst{2,4,5}}
\institute{
$^1$GSI Helmholtzzentrum f\"ur Schwerionenforschung GmbH, \\
64291 Darmstadt, Germany \\
\texttt{b.friman@gsi.de} \vspace{0.2cm} \\
$^2$TRIUMF, 4004 Wesbrook Mall, Vancouver, BC, V6T 2A3, Canada
\vspace{0.2cm} \\
$^3$Department of Physics, The Ohio State University, \\
Columbus, OH 43210, USA \\
\texttt{hebeler.4@osu.edu} \vspace{0.2cm} \\
$^4$ExtreMe Matter Institute EMMI, \\
GSI Helmholtzzentrum f\"ur Schwerionenforschung GmbH, \\
64291 Darmstadt, Germany \vspace{0.2cm} \\
$^5$Institut f\"ur Kernphysik, Technische Universit\"at Darmstadt, \\
64289 Darmstadt, Germany \\
\texttt{schwenk@physik.tu-darmstadt.de}}

\maketitle

\begin{abstract}
We discuss renormalization group approaches to strongly interacting
Fermi systems, in the context of Landau's theory of Fermi liquids
and functional methods, and their application to neutron matter.
\end{abstract}

\section{Introduction}

In these lecture notes we discuss developments using renormalization
group (RG) methods for strongly interacting Fermi systems and their
application to neutron matter. We rely on material from the review of
Shankar~\cite{Shankar}, the lecture notes by
Polchinski~\cite{Polchinski}, and work on the functional RG, discussed
in the lectures of Gies~\cite{Gies} and in the recent review by
Metzner {\it et al.}~\cite{Metzner}. The lecture notes are intended to
show the strengths and flexibility of the RG for nucleonic matter, and
to explain the ideas in more detail.

We start these notes with an introduction to Landau's theory of normal
Fermi liquids~\cite{Landau1,Landau2,Landau3}, which make the
concept of a quasiparticle very clear. Since Landau's work, this
concept has been successfully applied to a wide range of many-body
systems. In the quasiparticle approximation it is assumed that the
relevant part of the excitation spectrum of the one-body propagator
can be incorporated as an effective degree of freedom, a
quasiparticle. In Landau's theory of normal Fermi liquids this
assumption is well motivated, and the so-called background
contributions to the one-body propagator are included in the
low-energy couplings of the theory. In microscopic calculations and in
applications of the RG to many-body systems, the quasiparticle
approximation is physically motivated and widely used due to the great
reduction of the calculational effort.

\section{Fermi liquid theory}

\subsection{Basic ideas}
\label{basic}

Much of our understanding of strongly interacting Fermi systems at low
energies and temperatures goes back to the seminal work of Landau in
the late fifties~\cite{Landau1,Landau2,Landau3}. Landau was able to
express macroscopic observables in terms of microscopic properties of
the elementary excitations, the so-called quasiparticles, and their
residual interactions. In order to illustrate Landau's arguments here,
we consider a uniform system of non-relativistic spin-$1/2$ fermions
at zero temperature.

Landau assumed that the low-energy, elementary excitations of the
interacting system can be described by effective degrees of freedom,
the quasiparticles. Due to translational invariance, the states of the
uniform system are eigenstates of the momentum operator. The
quasiparticles are much like single-particle states in the sense that
for each momentum there is a well-defined quasiparticle energy. We
stress, however, that a quasiparticle state is not an energy
eigenstate, but rather a resonance with a non-zero width. For
quasiparticles close to the Fermi surface, the width is small and the
corresponding life-time is large; hence the quasiparticle concept is
useful for time scales short compared to the quasiparticle
life-time. Landau assumed that there is a one-to-one correspondence
between the quasiparticles and the single-particle states of a free
Fermi gas. For a superfluid system, this one-to-one correspondence
does not exist, and Landau's theory must be suitably modified, as
discussed by Larkin and Migdal~\cite{LarkinMigdal} and
Leggett~\cite{Leggett}.  Whether the quasiparticle concept is useful
for a particular system can be determined by comparison with
experiment or by microscopic calculations based on the underlying theory.

\begin{figure}[t]
\begin{center}
\includegraphics[scale=0.9,clip=]{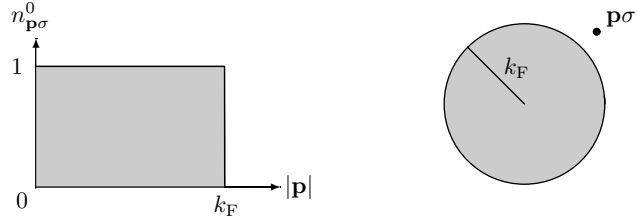}
\end{center}
\caption{Zero-temperature distribution function of a free Fermi gas
in the ground state (left) and with one added particle (right).}
\label{freegas}
\end{figure}

The one-to-one correspondence starts from a free Fermi gas consisting
of $N$ particles, where the ground state is given by a filled Fermi
sphere in momentum space, see Fig.~\ref{freegas}. The particle number
density $n$ and the ground-state energy $E_0$ are given by (with
$\hbar=c=1$)
\begin{equation}
n = \frac{1}{V} \sum_{\mathbf{p} \sigma} n_{\mathbf{p} \sigma}^0 
= \frac{\kf^3}{3 \pi^2} \quad \text{and} \quad
E_0 = \sum_{\mathbf{p} \sigma} 
\frac{\mathbf{p}^2}{2 m} \, n_{\mathbf{p} \sigma}^0
= \frac{3}{5} \frac{\kf^2}{2m} \, N \,,
\end{equation}
where $\kf$ denotes the Fermi momentum, $V$ the volume, and
$n_{\mathbf{p} \sigma}^0 = \theta(\kf - |\mathbf{p}|)$ is the Fermi-Dirac
distribution function at zero temperature for particles with momentum
$\mathbf{p}$, spin projection $\sigma$, and mass $m$.
By adding particles or holes, the distribution function is changed by 
$\delta n_{\mathbf{p}\sigma} = n_{\mathbf{p}\sigma}-n_{\mathbf{p}\sigma}^0$, 
and the total energy of the system by
\begin{equation}
\delta E = E - E_0 = \sum_{\mathbf{p} \sigma} \frac{\mathbf{p}^2}{2m} \, 
\delta n_{\mathbf{p} \sigma} \,.
\end{equation}
When a particle is added in the state $\mathbf{p}\sigma$, one has
$\delta n_{\mathbf{p}\sigma} =1$ and when a particle is removed 
(a hole is added) $\delta n_{\mathbf{p}\sigma} =-1$.

In the interacting system the corresponding state is one with a
quasiparticle added or removed, and the change in energy is given by
\begin{equation}
\delta E = \sum_{\mathbf{p} \sigma} \varepsilon_{\mathbf{p} \sigma} \,
\delta n_{\mathbf{p} \sigma} \,,
\end{equation}
where $\varepsilon_{\mathbf{p} \sigma}=\delta E/\delta
n_{\mathbf{p}\sigma}$ denotes the quasiparticle energy. When two or
more quasiparticles are added to the system, an additional term takes
into account the interaction between the quasiparticles:
\begin{equation}
\delta E = \sum_{\mathbf{p} \sigma} \varepsilon^0_{\mathbf{p} \sigma} \, 
\delta n_{\mathbf{p} \sigma} + \frac{1}{2V} 
\sum_{\mathbf{p}_1 \sigma_1, \mathbf{p}_2 \sigma_2} 
f_{\mathbf{p}_1 \sigma_1 \mathbf{p}_2 \sigma_2} \, \delta n_{\mathbf{p}_1 \sigma_1} \, 
\delta n_{\mathbf{p}_2 \sigma_2} \,.
\label{eq:delta_E}
\end{equation}
Here $\varepsilon^0_{\mathbf{p} \sigma}$ is the quasiparticle energy
in the ground state. In the next section, we will show that the
expansion in $\delta n$ is general and does not require weak
interactions. The small expansion parameter in Fermi liquid theory is
the density of quasiparticles, or equivalently the excitation energy,
and not the strength of the interaction. This allows a systematic
treatment of strongly interacting systems at low temperatures.

The second term in Eq.~(\ref{eq:delta_E}), the quasiparticle
interaction $f_{\mathbf{p}_1 \sigma_1 \mathbf{p}_2 \sigma_2}$, has no
correspondence in the non-interacting Fermi gas. In an
excited state with more than one quasiparticle, the quasiparticle
energy is modified according to
\begin{equation}
\varepsilon_{\mathbf{p} \sigma}
= \frac{\delta E}{\delta n_{\mathbf{p}\sigma}}
= \varepsilon^0_{\mathbf{p} \sigma}+\frac{1}{V}
\sum_{\mathbf{p}_2 \sigma_2} f_{\mathbf{p} \sigma \mathbf{p}_2 \sigma_2} \, \delta 
n_{\mathbf{p}_2 \sigma_2} \,,
\label{eq:qp-energy-var}
\end{equation}
where the changes are effectively proportional to the quasiparticle density.

The quasiparticle interaction can be understood microscopically from
the second variation of the energy with respect to the quasiparticle
distribution,
\begin{equation}
f_{\mathbf{p}_1 \sigma_1 \mathbf{p}_2 \sigma_2} = V \, 
\frac{\delta^2 E}{\delta n_{\mathbf{p}_1 \sigma_1} \delta n_{\mathbf{p}_2 \sigma_2}} 
= V \, \frac{\delta \varepsilon_{\mathbf{p}_1 \sigma_1}}{\delta 
n_{\mathbf{p}_2 \sigma_2}} \,.
\label{eq:f}
\end{equation}
As discussed in detail in Section~\ref{functional}, this variation
diagrammatically corresponds to cutting one of the fermion lines in a
given energy diagram and labeling the incoming and outgoing fermion by
$\mathbf{p}_1 \sigma_1$, followed by a second variation leading to
$\mathbf{p}_2 \sigma_2$. For the uniform system, the resulting
contributions to $f_{\mathbf{p}_1 \sigma_1 \mathbf{p}_2 \sigma_2}$ are
quasiparticle reducible in the particle-particle and in the exchange
particle-hole (induced interaction) channels, but irreducible in the
direct particle-hole (zero sound) channel. The zero-sound reducible
diagrams are generated by the particle-hole scattering equation.

In normal Fermi systems, the quasiparticle concept makes sense only
for states close to the Fermi surface, where the quasiparticle
life-time $\tau_{\mathbf{p}}$ is long. The leading term is quadratic in
the momentum difference from the Fermi surface~\cite{BaymPethick},
$1/\tau_{\mathbf{p}} \sim (p - \kf)^2$, while the dependence of the
quasiparticle energy is linear, $\varepsilon_{\mathbf{p}} - \mu \sim
(p - \kf)$.  Thus, the condition
\begin{equation}
|\varepsilon_{\mathbf{p}} - \mu | \gg \frac{1}{\tau_{\mathbf{p}}} \,,
\end{equation}
which is needed for the quasiparticle to be well defined, is satisfied
by states close enough to the Fermi surface. Generally, quasiparticles
are useful for time scales $\tau \ll \tau_{\mathbf{p}}$ and thus for
high frequencies $|\omega| \tau_{\mathbf{p}} \gg 1$.  In particular,
states deep in the Fermi sea, which are occupied in the ground-state
distribution, do not correspond to well-defined
quasiparticles. Accordingly, we refer to the interacting ground state
that corresponds to a filled Fermi sea in the non-interacting system
as a state with no quasiparticles. In a weakly excited state the
quasiparticle distribution $\delta n_{\mathbf{p} \sigma}$ is generally
non-zero only for states close to the Fermi surface.

For low-lying excitations, the quasiparticle energy
$\varepsilon_{\mathbf{p}\sigma}$ and interaction $f_{\mathbf{p}_1
\sigma_1 \mathbf{p}_2 \sigma_2}$ is needed only for momenta close to
the Fermi momentum $\kf$. It is then sufficient to retain the leading
term in the expansion of $\varepsilon_{\mathbf{p} \sigma} - \mu$
around the Fermi surface, and to take the magnitude of the
quasiparticle momenta in $f_{\mathbf{p}_1 \sigma_1 \mathbf{p}_2
\sigma_2}$ equal to the Fermi momentum. In an isotropic and
spin-saturated system ($N_{\uparrow} = N_{\downarrow}$), and if the
interaction between free particles is invariant under $SU(2)$ spin
symmetry (so that there are no non-central contributions, such 
as $\sim {\bm \sigma} \cdot \mathbf{p}$ to the energy), we have
\begin{equation}
\varepsilon_{\mathbf{p} \sigma} - \mu = \varepsilon_p - \mu \approx
\vf (p - \kf) + \ldots \,,
\end{equation}
where $\vf = \kf/m^*$ denotes the Fermi velocity and $m^*$ is the
effective mass. In addition, the quasiparticle interaction can be 
decomposed as
\begin{equation}
f_{\mathbf{p}_1 \sigma_1 \mathbf{p}_2 \sigma_2} = f_{\mathbf{p}_1 \mathbf{p}_2}^s 
+ f^a_{\mathbf{p}_1 \mathbf{p}_2} \: \spin \,,
\label{eq:spin}
\end{equation}
where
\begin{equation}
f^s_{\mathbf{p}_1 \mathbf{p}_2} = \frac{1}{2} \, \bigl(
f_{\mathbf{p}_1 \uparrow \mathbf{p}_2 \uparrow} 
+ f_{\mathbf{p}_1 \uparrow \mathbf{p}_2 \downarrow} \bigr)
\quad \text{and} \quad
f^a_{\mathbf{p}_1 \mathbf{p}_2} = \frac{1}{2} \,
\bigl( f_{\mathbf{p}_1 \uparrow \mathbf{p}_2 \uparrow} 
- f_{\mathbf{p}_1 \uparrow \mathbf{p}_2 \downarrow} \bigr) \,.
\end{equation}
In nuclear physics the notation $f_{\mathbf{p}_1 \mathbf{p}_2} =
f^s_{\mathbf{p}_1 \mathbf{p}_2}$ and $g_{\mathbf{p}_1 \mathbf{p}_2} =
f^a_{\mathbf{p}_1 \mathbf{p}_2}$ is generally used, and the
quasiparticle interaction includes additional terms that take into
account the isospin dependence and non-central tensor 
contributions~\cite{MigdalBook,GerryPR,tensor}. However, for our
discussion here, the spin and isospin dependence is not important.

For the uniform system, Eq.~(\ref{eq:f}) yields the quasiparticle
interaction only for forward scattering (low momentum transfers). In
the particle-hole channel, this corresponds to the long-wavelength
limit. This restriction, which is consistent with considering low
excitation energies, constrains the momenta $\mathbf{p}_1$ and
$\mathbf{p}_2$ to be close to the Fermi surface, $|\mathbf{p}_1| =
|\mathbf{p}_2| = \kf$. The quasiparticle interaction then depends only
on the angle between $\mathbf{p}_1$ and $\mathbf{p}_2$. It is
convenient to expand this dependence on Legendre polynomials
\begin{equation}
f_{\mathbf{p}_1 \mathbf{p}_2}^{s/a} = f^{s/a}(\cos \theta_{\mathbf{p}_1 \mathbf{p}_2}) 
= \sum_l \, f_l^{s/a} \, P_l(\cos \theta_{\mathbf{p}_1 \mathbf{p}_2}) \,,
\label{eq:f_expansion}
\end{equation}
and to define the dimensionless Landau Parameters $F_l^{s/a}$ by
\begin{equation}
F_l^{s/a} = N(0) \, f_l^{s/a} \,,
\end{equation}
where $N(0) = \frac{1}{V} \sum_{\mathbf{p} \sigma} 
\delta(\varepsilon_{\mathbf{p} \sigma} - \mu) = m^* \kf/\pi^2$
denotes the quasiparticle density of states at the Fermi surface.

The Landau parameters can be directly related to macroscopic
properties of the system. $F_1^s$ determines the effective mass
and the specific heat $c_V$,
\begin{align}
\frac{m^*}{m} &= 1 + \frac{F^s_1}{3} \,, \\
c_V &= \frac{m^* \kf}{3} \, k_B^2 T \,,
\end{align}
while the compressibility $K$ and incompressibility $\kappa$ are given
by $F_0^s$,
\begin{align}
K &= - \frac{1}{V} \frac{\partial V}{\partial P} = \frac{1}{n^2} 
\frac{\partial n}{\partial \mu} = \frac{1}{n^2} \frac{N(0)}{1 + F_0^s} \,, \\
\kappa &= \frac{9}{n K}= - \frac{9 V}{n} \frac{\partial P}{\partial V} 
= 9 \, \frac{\partial P}{\partial n} = \frac{3 \kf^2}{m^*} \, (1 + F_0^s) \,.
\end{align}
Moreover, the spin susceptibility $\chi_m$ is related to $F_0^a$,
\begin{equation}
\chi_m = \frac{\partial m}{\partial H} = \beta^2 \frac{N(0)}{1+F_0^a} \,,
\end{equation}
for spin-$1/2$ fermions with magnetic moment $\beta = ge/(4m)$
and gyromagnetic ratio $g$. Finally,
a stability analysis of the Fermi surface against small 
amplitude deformations leads to the Pomeranchuk criteria~\cite{Pomeranchuk}
\begin{equation}
F_l^{s/a} > - (2l+1) \,.
\end{equation}
For instance $F_0^{s/a} < -1$ implies an instability against 
spontaneous growth of density/spin fluctuations.

Landau's theory of normal Fermi liquids is an effective low-energy
theory in the modern sense~\cite{Shankar,Polchinski}. The effective
theory incorporates the symmetries of the system and the low-energy
couplings can be fixed by experiment or calculated microscopically
based on the underlying theory. Fermi Liquid theory has been very
successful in describing low-temperature Fermi liquids, in particular
liquid $^3$He. Applications to the normal phase are reviewed, for
example, in Baym and Pethick~\cite{BaymPethick} and Pines and
Nozi\`eres~\cite{PinesNozieres}, while we refer to W\"olfle and
Vollhardt~\cite{WoelfleVollhardt} for a description of the superfluid
phases. The first applications to nuclear systems were pioneered by
Migdal~\cite{MigdalBook} and first microscopic calculations for nuclei
and nuclear matter by Brown {\it et al.}~\cite{GerryPR}. Recently,
advances using RG methods for nuclear forces~\cite{Vlowk} have lead to
the development of a non-perturbative RG approach for nucleonic
matter~\cite{RGnm}, to a first complete study of the spin structure of
induced interactions~\cite{tensor}, and to new calculations of Fermi
liquid parameters~\cite{indint,Kaiser}.

\subsection{Three-quasiparticle interactions}

In Section~\ref{basic}, we introduced Fermi liquid theory as an
expansion in the density of quasiparticles $\delta n/V$. In
applications of Fermi liquid theory to date, even for liquid $^3$He,
which is a very dense and strongly interacting system, this expansion
is truncated after the second-order $(\delta n)^2$ term, including
only pairwise interactions of quasiparticles (see
Eq.~(\ref{eq:delta_E})). However, for a strongly interacting system,
there is a priori no reason that three-body (or higher-body)
interactions between quasiparticles are small. In this section, we
discuss the convergence of this expansion.  Three-quasiparticle
interactions arise from iterated two-body forces, leading to three-
and higher-body clusters in the linked-cluster expansion, or through
many-body forces. While three-body forces play an important role in
nuclear physics~\cite{RMP,matter,nuclei}, little is known about them
in other Fermi liquids. Nevertheless, in strongly interacting systems,
the contributions of many-body clusters can in general be significant,
leading to potentially important $(\delta n)^3$ terms in the Fermi
liquid expansion, also in the absence of three-body forces:
\begin{equation}
\delta E = \sum_1 \varepsilon^0_1 \: \delta n_1
+ \frac{1}{2V} \sum_{1,2} f^{(2)}_{1,2} \: \delta n_1 \, \delta n_2
+ \frac{1}{6V^2} \sum_{1,2,3} f^{(3)}_{1,2,3} \: \delta n_1 \, \delta n_2
\, \delta n_3 \,.
\label{eq:delta_E3}
\end{equation}
Here $f^{(n)}_{1,\ldots,n}$ denotes the $n$-quasiparticle interaction
(the Landau interaction is $f \equiv f^{(2)}$) and we have introduced
the short-hand notation $n \equiv {\bf p}_n \sigma_n$.

In order to better understand the expansion, Eq.~(\ref{eq:delta_E3}),
around the interacting ground state with $N$ fermions, consider
exciting or adding $N_q$ quasiparticles with $N_q \ll N$. The
microscopic contributions from many-body clusters or from many-body
forces can be grouped into diagrams containing zero, one, two, three,
or more quasiparticle lines.  The terms with zero quasiparticle lines
contribute to the interacting ground state for $\delta n = 0$, whereas
the terms with one, two, and three quasiparticle lines contribute to
$\varepsilon^0_1$, $f^{(2)}_{1,2}$, and $f^{(3)}_{1,2,3}$,
respectively (these also depend on the ground-state density due to the
$N$ fermion lines). The terms with more than three quasiparticle lines
would contribute to higher-quasiparticle interactions. Because a
quasiparticle line replaces a line summed over $N$ fermions when going from
$\varepsilon^0_1$ to $f^{(2)}_{1,2}$, and from $f^{(2)}_{1,2}$ to
$f^{(3)}_{1,2,3}$, it is intuitively clear that the contributions due
to three-quasiparticle interactions are suppressed by $N_q/N$ compared
to two-quasiparticle interactions, and that the Fermi liquid expansion
is effectively an expansion in $N_q/N$ or
$n_q/n$~\cite{PinesNozieres}. 

Fermi liquid theory applies to normal Fermi systems at low energies
and temperatures, or equivalently at low quasiparticle densities. We
first consider excitations that conserve the net number of quasiparticles,
$\delta N = \sum_{\mathbf{p} \sigma} \delta n_{\mathbf{p} \sigma}=0$,
so that the number of quasiparticles equals the number of
quasiholes. This corresponds to the lowest energy excitations of
normal Fermi liquids. We denote their energy scale by
$\Delta$. Excitations with one valence particle or quasiparticle added
start from energies of order the chemical potential $\mu$. In the case
of $\delta N=0$, the contributions of two-quasiparticle interactions
are of the same order as the first-order $\delta n$ term, but
three-quasiparticle interactions are suppressed by
$\Delta/\mu$~\cite{Chris}. This is the reason that Fermi liquid theory
with only two-body Landau parameters is so successful in describing
even strongly interacting and dense Fermi liquids. This counting is
best seen from the variation of the free energy $F=E-\mu N$,
\begin{align}
\delta F &= \delta(E-\mu N) \nonumber \\
&= \sum_1 (\varepsilon^0_1 - \mu) \, \delta n_1
+ \frac{1}{2V} \sum_{1,2} f^{(2)}_{1,2} \: \delta n_1 \, \delta n_2
+ \frac{1}{6V^2} \sum_{1,2,3} f^{(3)}_{1,2,3} \: \delta n_1 \, \delta n_2
\, \delta n_3 \,,
\label{eq:delta_F3}
\end{align}
which for $\delta N=0$ is equivalent to $\delta E$ of
Eq.~(\ref{eq:delta_E3}). The quasiparticle distribution
is $|\delta n_{\mathbf{p} \sigma}| \sim 1$ within a shell around the
Fermi surface $|\varepsilon^0_{\mathbf{p} \sigma} - \mu| \sim
\Delta$. The first-order $\delta n$ term is therefore proportional to
$\Delta$ times the number of quasiparticles
$\sum_{\mathbf{p} \sigma} |\delta n_{\mathbf{p} \sigma}| = N_q
\sim N (\Delta/\mu)$, and
\begin{equation}
\sum_1 (\varepsilon^0_1 - \mu) \, \delta n_1 \sim \frac{N
\Delta^2}{\mu} \,.
\end{equation}
Correspondingly, the contribution of two-quasiparticle interactions
yields
\begin{equation}
\frac{1}{2V} \sum_{1,2} f^{(2)}_{1,2} \: \delta n_1 \, \delta n_2
\sim \frac{1}{V} \, \langle f^{(2)} \rangle \, \biggl(\frac{N 
\Delta}{\mu}\biggr)^2 
\sim \langle F^{(2)} \rangle \, \frac{N \Delta^2}{\mu} \,,
\label{eq:f2scaling}
\end{equation}
where $\langle F^{(2)} \rangle = n \, \langle f^{(2)} \rangle /\mu$ is
an average dimensionless coupling on the order of the Landau
parameters. Even in the strongly interacting, scale-invariant case
$\langle f^{(2)} \rangle \sim 1/\kf$;
hence $\langle F^{(2)} \rangle \sim 1$ and the contribution of
two-quasiparticle interactions is of the same order as the first-order
term. However, the three-quasiparticle contribution is of order
\begin{equation}
\frac{1}{6V^2} \sum_{1,2,3} f^{(3)}_{1,2,3} \: \delta n_1 \, \delta n_2
\, \delta n_3 
\sim \frac{n^2}{\mu} \, \langle f^{(3)} \rangle \, \frac{N \Delta^3}{\mu^2}
\sim \langle F^{(3)} \rangle \,\frac{N \Delta^3}{\mu^2} \,.
\label{eq:f3scaling}
\end{equation}
Therefore at low excitation energies this is suppressed by
$\Delta/\mu$, compared to two-quasiparticle interactions, even if the
dimensionless three-quasiparticle interaction $\langle F^{(3)} \rangle
= n^2 \langle f^{(3)} \rangle /\mu$ is strong (of order 1). Similarly,
higher $n$-body interactions are suppressed by
$(\Delta/\mu)^{n-2}$. Normal Fermi systems at low energies are weakly
coupled in this sense.  The small parameter is the ratio of the
excitation energy per particle to the chemical potential. These
considerations hold for all normal Fermi systems where the
underlying interparticle interactions are finite range.

The Fermi liquid expansion in $\Delta/\mu$ is equivalent to an
expansion in $N_q/N \sim \Delta/\mu$, the ratio of the number of
quasiparticles and quasiholes $N_q$ to the number of particles $N$ in
the interacting ground state, or an expansion in the density of
excited quasiparticles over the ground-state density, $n_q/n$. For
the case where $N_q$ quasiparticles or valence particles are added to
a Fermi-liquid ground state, $\delta N \neq 0$ and the first-order
term is
\begin{equation}
\sum_1 \varepsilon^0_1 \: \delta n_1 \sim \mu N_q 
\sim \mu \, \frac{N \Delta}{\mu}
\sim N \Delta \,,
\end{equation}
while the contribution of two-quasiparticle interactions is suppressed
by $N_q/N \sim \Delta/\mu$ and that of three-quasiparticle
interactions by $(N_q/N)^2$.

Therefore, either for $\delta N=0$ or $\delta N \neq 0$, the
contributions of three-quasiparticle interactions to normal Fermi
systems at low excitation energies are suppressed by the ratio of the
quasiparticle density to the ground-state density, or equivalently by
the ratio of the excitation energy over the chemical potential.  This
holds for excitations that conserve the number of particles (excited
states of the interacting ground state) as well as for excitations
that add or remove particles. This suppression is general and applies
to strongly interacting systems even with strong, but finite-range
three-body forces. However, this does not imply that the contributions
from three-body forces to the interacting ground-state energy (the
energy of the core nucleus in the context of shell-model
calculations), to quasiparticle energies, or to two-quasiparticle
interactions are small. The argument only applies to the effects of
residual three-body interactions at low energies.

\subsection{Microscopic foundation of Fermi liquid theory}
\label{microscopic}

A central object in microscopic approaches to many-body systems is the 
one-body (time-ordered) propagator or Green's function $G$ defined by
\begin{equation}
G(1,2) = -i \, \langle 0 | {\mathcal T} \psi(1) \psi^\dagger(2) | 0
\rangle \,,
\label{Green}
\end{equation}
where $| 0 \rangle$ denotes the ground state of the system, ${\mathcal
T}$ is the time-ordering operator, $\psi$ and $\psi^\dagger$
annihilate and create a fermion, respectively, and 1, 2 are short hand
for space, time and internal degrees of freedom (such as spin and
isospin).  For a translationally invariant spin-saturated system that
is also invariant under rotations in spin space, the Green's function
is diagonal in spin and can be written in momentum space as
\be
G(\omega,\mathbf{p}) \, \delta_{\sigma_1 \sigma_2}
= \int d(1-2) \, G(1,2) \, e^{i \omega (t_1 - t_2) - i \mathbf{p} \cdot
(\mathbf{x}_1 - \mathbf{x}_2)}
= \frac{\delta_{\sigma_1 \sigma_2}}{\omega - \frac{p^2}{2 m} 
- \Sigma(\omega,\mathbf{p})} \,,
\ee
where $\Sigma(\omega,\mathbf{p})$ defines the self-energy. For an
introduction to many-body theory and additional details, we refer
to the books by Fetter and Walecka~\cite{FW}, Abrikosov, Gor'kov
and Dzyaloshinski~\cite{AGD}, Negele and Orland~\cite{NO}, and
Altland and Simons~\cite{AS}.

\begin{figure}[t]
\begin{center}
\includegraphics[scale=0.8,clip=]{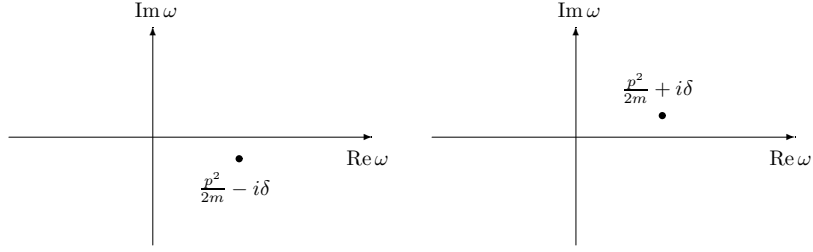}
\end{center}
\caption{Analytic structure of the free one-body Green's function $G_0$
in the complex $\omega$ plane with simple poles for $p>\kf$ (left) and 
$p<\kf$ (right).}
\label{Gpoles}
\end{figure}

Without interactions the self-energy vanishes and consequently the free 
Green's function $G_0$ reads
\be
G_0 (\omega,\mathbf{p}) = \frac{1}{\omega - \frac{p^2}{2 m} + i 
\delta_\mathbf{p}} = 
\frac{1 - n^0_{\mathbf{p}}}{\omega - \frac{p^2}{2m} + i \delta} 
+ \frac{n^0_{\mathbf{p}}}{\omega - \frac{p^2}{2m} - i \delta} \,,
\ee
where $\delta_\mathbf{p} = \delta \: {\rm sign}(p - \kf)$ and $\delta$
is a positive infinitesimal. The free Green's function has simple poles,
as illustrated in Fig.~\ref{Gpoles}, and the imaginary part takes the
form
\begin{equation}
{\rm Im} \, G_0(\omega,\mathbf{p}) = - \pi \, 
(1 - 2 n^0_{\mathbf{p}}) \, \delta \biggl( \omega - \frac{p^2}{2m} 
\biggr) \,.
\end{equation}
The single-particle spectral function $\rho (\omega,\mathbf{p})$ is
determined by the imaginary part of the retarded propagator
\begin{align}
G^R(1,2) &= -i \, \theta(t_1-t_2) \,
\langle 0 | \bigl\{ \psi(1) , \psi^\dagger(2) \bigr\} | 0 \rangle \,, \\
\rho (\omega,\mathbf{p}) &= - \frac{1}{\pi} \, 
{\rm Im} \, G^R(\omega,\mathbf{p}) \,,
\label{rho_def}
\end{align}
where $\{\:,\:\}$ denotes the anticommutator. The retarded propagator
is analytic in the upper complex $\omega$ plane and fulfills
Kramers-Kronig relations, which relate the real and imaginary
parts. Physically this implies that all modes are propagating forward
in time and causality is fulfilled. Therefore, response functions are
usually expressed in terms of the retarded propagator.

In a non-interacting system the retarded propagator is given by
\begin{align}
G_0^R(\omega,\mathbf{p}) &= \frac{1}{\omega - \frac{p^2}{2m} 
+ i \delta} \,, \\
{\rm Im} \, G_0^R(\omega,\mathbf{p}) &= - \pi \, \delta 
\biggl( \omega - \frac{p^2}{2m} \biggr) \,,
\end{align}
which implies that the free spectral function is a delta
function $\rho_0(\omega,\mathbf{p}) = \delta(\omega - \frac{p^2}{2m})$.
This simple form follows from the fact that single-particle 
plane-wave states are eigenstates of the non-interacting Hamiltonian.

In the interacting case the situation is more complicated. Here the 
quasiparticle energy is given implicitly by the Dyson equation
\begin{equation}
\varepsilon_{\mathbf{p}} = \frac{p^2}{2m} + \Sigma( 
\varepsilon_{\mathbf{p}},\mathbf{p}) \,.
\label{eq:dyson-qp-energy}
\end{equation}
At the chemical potential $\omega = \mu$,
the imaginary part of the self-energy vanishes,
\begin{equation}
{\rm Im} \, \Sigma(\mu,\mathbf{p}) = 0 \,,
\end{equation}
and the quasiparticle life-time $\tau_{\mathbf{p}} \to \infty$
for $|\mathbf{p}| \to \kf$.%
\footnote{At non-zero temperature, the imaginary part of the self-energy 
never vanishes and the quasiparticle life-time is finite. However, 
for $T \ll \mu$ and $\omega \approx \mu$, the life-time is large and the 
quasiparticle concept is useful.}  For $\omega \neq \mu$, the imaginary 
part of the self-energy obeys
\begin{align}
{\rm Im} \, \Sigma(\omega,\mathbf{p}) &< 0 \,\text{, for } \omega > \mu \,,\\
{\rm Im} \, \Sigma(\omega,\mathbf{p}) &> 0 \,\text{, for } \omega < \mu \,.
\end{align}
The retarded self-energy, which enters the retarded Green's function 
\begin{equation}
G^R (\omega,\mathbf{p}) = \frac{1}{\omega - \frac{p^2}{2m} - 
\Sigma^R(\omega,\mathbf{p})} \,,
\end{equation}
is related to the time-ordered one through
\begin{align}
{\rm Re} \, \Sigma^R(\omega,\mathbf{p}) &= {\rm Re} \,
\Sigma(\omega,\mathbf{p}) \,, \\[1mm]
{\rm Im} \, \Sigma^R(\omega,\mathbf{p}) &= \biggl\{
\begin{array}{l} + {\rm Im} \, \Sigma(\omega,\mathbf{p}) < 0 \,
\text{, for } \omega > \mu \,, \\[1mm] 
- {\rm Im} \, \Sigma(\omega,\mathbf{p}) < 0 \,
\text{, for } \omega < \mu \,. \end{array}
\end{align}
Using Eq.~(\ref{rho_def}), one finds the general form of the spectral function 
\begin{equation}
\rho(\omega,\mathbf{p}) = - \frac{1}{\pi}
\frac{{\rm Im} \, \Sigma^R(\omega,\mathbf{p})}
{\bigl[\omega - \frac{p^2}{2m} - {\rm Re} \, \Sigma^R(\omega,\mathbf{p})
\bigr]^2 + \bigl[{\rm Im} \, \Sigma^R(\omega,\mathbf{p})\bigr]^2} \,.
\end{equation}
In the interacting case the single-particle strength is therefore, for 
a given momentum state, fragmented in energy, and the spectral function
provides a measure of the single-particle strength in the eigenstates 
of the Hamiltonian, with the normalization $\int_{-\infty}^{\infty} d\omega \,
\rho(\omega,\mathbf{p}) = 1$.

\begin{figure}[t]
\begin{center}
\includegraphics[scale=0.8,clip=]{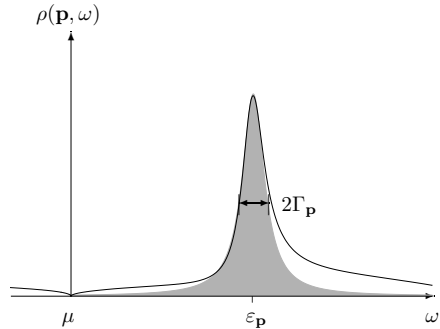}
\end{center}
\caption{Spectral function $\rho(\omega,\mathbf{p})$ for a given momentum
$p \gtrsim \kf$ as a function of frequency $\omega$. The shaded part marks the
quasiparticle peak, with quasiparticle width $\Gamma_{\mathbf{p}}$ and
single-particle strength $z_{\mathbf{p}}$.}
\label{fig:quasiparticle}
\end{figure}

The spectral or K\"all\'en-Lehmann representation of the Green's function,
\begin{equation}
G^R(\omega,\mathbf{p}) = \int_{-\infty}^{\infty} d\omega' \: 
\frac{\rho(\omega',\mathbf{p})}{\omega - \omega' + i \delta} \,,
\label{kallen}
\end{equation}
follows from analyticity and implies that the full propagator is
completely determined by the spectral function. Using
Eq.~(\ref{kallen}) and the normalization condition, the asymptotic
($|\omega| \rightarrow \infty$) behavior of both
$G^R(\omega,\mathbf{p})$ and $G(\omega,\mathbf{p}) \sim 1/\omega$
follows. Furthermore, the singularities of the full Green's function (that
correspond to eigenvalues of the Hamiltonian) are all located on the
real axis and result in a cut along the real axis in the continuum
limit. The quasiparticle pole, on the other hand, is located off the
real axis, on an unphysical Riemann sheet,\footnote{%
This is readily seen by evaluating $G(z=\omega\pm i\delta,\mathbf{p})$
for a spectral function of the quasiparticle form,
\be
\rho(\omega,\mathbf{p}) = \frac{1}{\pi}
\frac{\Gamma_\mathbf{p}/2}{(\omega - \varepsilon_\mathbf{p})^2 +
\Gamma_\mathbf{p}^2/4} \,, 
\ee
using the K\"all\'en-Lehmann representation for the Green's function
at a complex argument,
\be
G(z,\mathbf{p}) = \int_{-\infty}^{\infty} d\omega' \:
\frac{\rho(\omega',\mathbf{p})}{z - \omega'} \,.
\ee
One then finds
that $G(\omega\pm i\delta,\mathbf{p}) = (\omega -
\varepsilon_\mathbf{p} \pm i \Gamma_\mathbf{p}/2)^{-1}$. This
implies that for $\omega$ in the upper half plane the quasiparticle
pole is in the lower half plane and vice versa.}
with the distance
to the real axis given by the quasiparticle width,
\begin{align}
\text{quasiparticle pole $(p \gtrsim \kf)$: } \: \omega_\mathbf{p}^{\rm qp} &=
\varepsilon_{\mathbf{p}} - i \Gamma_{\mathbf{p}} \,, \nonumber \\
\text{quasihole pole $(p \lesssim \kf)$: } \: \omega_\mathbf{p}^{\rm qp} &=
\varepsilon_{\mathbf{p}} + i \Gamma_{\mathbf{p}} \,.
\end{align}
A pole close to the real axis gives rise to a peak in the 
single-particle strength, as illustrated in Fig.~\ref{fig:quasiparticle}.
Hence, microscopically a quasiparticle or quasihole is identified
by a well-defined peak in the spectral function. In other words, 
the excitation of a quasiparticle corresponds to the coherent 
excitation of several eigenstates $H | \psi_i \rangle = E_i | \psi_i
\rangle$ of the Hamiltonian, with similar energies $E_i$ ($\omega'
\approx \varepsilon_{\mathbf{p}}$ in Eq.~(\ref{kallen})) spread over 
the quasiparticle width $\Gamma_{\mathbf{p}}=\tau_{\mathbf{p}}^{-1}$.
A quasiparticle created at $t=0$ then propagates in time as
\be
| \psi_{\rm qp} (t) \rangle = \sum_i \, c_i \, e^{-i E_i t} \, | \psi_i 
\rangle \,.
\ee
For short times, $t \ll 1/\Gamma_{\mathbf{p}}$,
the eigenstates remain coherent and the quasiparticle is well defined, 
while for $t \gg 1/\Gamma_{\mathbf{p}}$, the phase coherence is lost and
the quasiparticle decays.

Using this definition of a quasiparticle, the full propagator can be
formally separated into a quasiparticle part and a smooth
background $\phi(\omega,\mathbf{p})$:
\begin{equation}
G(\omega,\mathbf{p}) = \frac{z_{\mathbf{p}}}{\omega - \varepsilon_{\mathbf{p}} 
+ i \Gamma_{\mathbf{p}}} + \phi(\omega,\mathbf{p}) \,,
\label{eq:G_qp}
\end{equation}
where the single-particle strength $z_{\mathbf{p}}$ carried by the 
quasiparticle is given by
\begin{equation}
z_{\mathbf{p}} = \biggl[ 1 - \frac{\partial \Sigma(\omega,\mathbf{p})}{
\partial \omega} \biggr|_{\omega = \varepsilon_{\mathbf{p}}} \biggr]^{-1} < 1 \,,
\label{zfact}
\end{equation}
and must be less than unity due to the normalization of the spectral 
function.

Close to the Fermi surface the width $\Gamma_{\mathbf{p}}$ is small, 
$\Gamma_{\mathbf{p}} \sim (p - \kf)^2$, and consequently quasiparticles
are well defined. For processes with a typical time scale $\tau < 
\tau_{\mathbf{p}} = \Gamma_{\mathbf {p}}^{-1}$, the contribution of the 
quasiparticle remains coherent, while that of the smooth background
is incoherent. Even for very small values of the single-particle 
strength $z_{\mathbf{p}}$, quasiparticles play a leading role at 
sufficiently low excitation energies.

\subsection{Scattering of quasiparticles}
\label{scattering}

\begin{figure}[t]
\begin{center}
\parbox{3.5cm}{\includegraphics[scale=0.78,clip=]{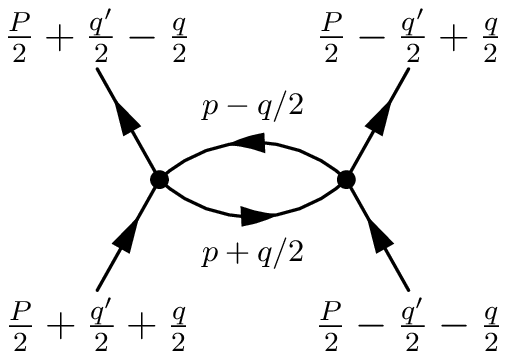}}
\hspace{6mm}
\parbox{3.5cm}{\includegraphics[scale=0.78,clip=]{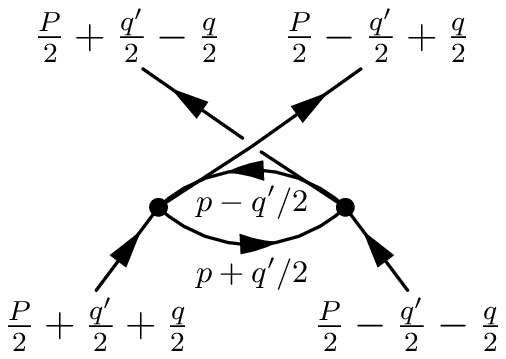}}
\hspace{4mm}
\parbox{2.6cm}{\includegraphics[scale=0.66,clip=]{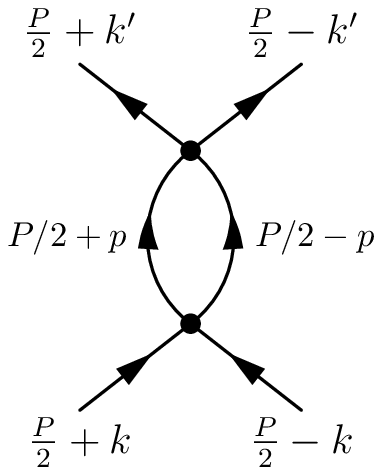}}
\end{center}
\caption{Scattering channels in a many-fermion system. The direct
particle-hole or zero-sound (ZS) channel (left), the exchange
particle-hole (ZS$'$) channel (middle) and the 
particle-particle/hole-hole
(BCS) channel (right). The relative incoming and outgoing 
four-momenta $k, k'$ are related to the momentum transfers 
$q, q'$ by $k=(q'+q)/2$ and $k'=(q'-q)/2$. The
center-of-mass momentum is denoted by $P$.\label{channels}}
\end{figure}

Quasiparticle scattering processes are in general described by the
Bethe-Salpeter equation. The quasiparticle scattering amplitude is
given by the full four-point function $\Gamma$, which includes
contributions from scattering in the channels shown in
Fig.~\ref{channels}.  For small $q = (\omega,\mathbf{q})$, the
contribution of the direct particle-hole or zero-sound (ZS) channel is
singular due to a pinching of the integration contour by the
quasiparticle poles of the two intermediate propagators for any
external momenta $p_1 = (P+q')/2$ and $p_2 = (P-q')/2$, as discussed
below and in detail in Ref.~\cite{AGD}.  In contrast, for small $q$
the exchange particle-hole (ZS$'$) and the particle-particle/hole-hole
(BCS) channels are smooth for almost all external
momenta.\footnote{The BCS singularity for back-to-back scattering,
$P=0$, is discussed in Section~\ref{RGnm}.}

\begin{figure}[t]
\begin{center}
\parbox{2.6cm}{\includegraphics[scale=0.82,clip=]{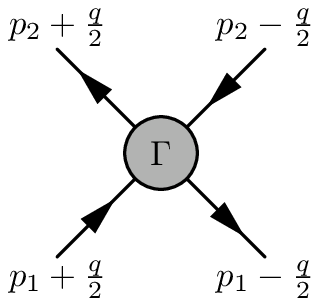}}
$\quad = \quad$
\parbox{2.6cm}{\includegraphics[scale=0.82,clip=]{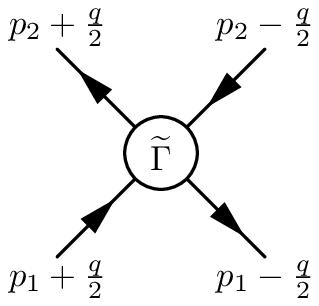}}
$\quad + \quad$
\parbox{2.6cm}{\includegraphics[scale=0.82,clip=]{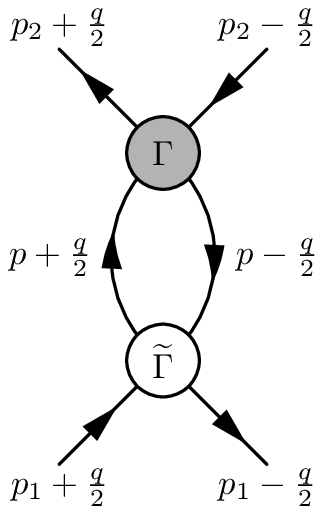}}
\end{center}
\caption{The Bethe-Salpeter equation in the ZS channel, Eq.~(\ref{BSZSeq}).
\label{BSZS}}
\end{figure}

The Bethe-Salpeter equation that sums all ZS-channel reducible diagrams
is shown diagrammatically in Fig.~\ref{BSZS} and reads
\begin{multline}
\Gamma(p_1, p_2; q) = \widetilde{\Gamma}(p_1, p_2; q) \\
-i \int \frac{d^4 p}{(2 \pi)^4} \, \widetilde{\Gamma}(p_1, p; q) \,
G(p + q/2) \, G(p - q/2) \, \Gamma(p, p_2; q) \,,
\label{BSZSeq}
\end{multline}
where $\widetilde{\Gamma}$ denotes the ZS-channel irreducible
four-point function and we suppress the spin of the fermions for
simplicity. The singular part of the two intermediate propagators is
obtained by using the quasiparticle representation of the Green's
function, Eq.~(\ref{eq:G_qp}), and for $q \to 0$ one finds
\begin{align}
&G(p+q/2) \, G(p-q/2) =
\frac{z_{\mathbf{p}+\mathbf{q}/2}}{\varepsilon + \omega/2 - \mu - \vf 
(|\mathbf{p}+\mathbf{q}|/2 - \kf) + i \delta_{|\mathbf{p}+\mathbf{q}|/2}} 
\nonumber \\
&\times \frac{z_{\mathbf{p}-\mathbf{q}/2}}{\varepsilon - \omega/2 - \mu - \vf 
(|\mathbf{p}-\mathbf{q}|/2 - \kf) + i \delta_{|\mathbf{p}-\mathbf{q}|/2}} + 
\phi_2(p) \,, \nonumber \\
&= 2 \pi i \, z_{\kf}^2 \, \frac{|\mathbf{q}| 
\cos\theta_{\mathbf{p}\mathbf{q}}}{
\omega - \vf |\mathbf{q}| \cos\theta_{\mathbf{p}\mathbf{q}}} \,
\delta(\varepsilon - \mu) \, \delta(|\mathbf{p}| - \kf)
+\phi_2(p) \,,
\label{eq:GG}
\end{align}
where $p=(\varepsilon, \mathbf{p})$. The first term in Eq.~(\ref{eq:GG})
is the quasiparticle-quasihole part, which constrains the intermediate
states to the Fermi surface, and the contribution $\phi_2(p)$
includes at least one power of the smooth background $\phi(p)$. In 
addition, we have taken $q \to 0$ in all nonsingular terms and
neglected the quasiparticle width, which is
small close to the Fermi surface.

We observe that the quasiparticle-quasihole part of the particle-hole
propagator vanishes in the limit $|\mathbf{q}|/\omega \to 0$. Therefore,
we define
\begin{align}
\Gamma^{\omega}(p_1, p_2) &= \underset{\omega \rightarrow 0}{{\rm lim}} \bigl(
\Gamma (p_1, p_2; q) \bigr|_{|\mathbf{q}|=0} \bigr) \,, \\
\Gamma^{q}(p_1, p_2) &= \underset{|\mathbf{q}| \rightarrow 0}{{\rm lim}} \bigl(
\Gamma (p_1, p_2; q) \bigr|_{\omega=0} \bigr) \,.
\end{align}
Using the quasiparticle-quasihole representation of the particle-hole
propagator, Eq.~(\ref{eq:GG}), the Bethe-Salpeter equation in the ZS
channel takes the form
\begin{multline}
\Gamma(p_1,p_2; q) = \widetilde{\Gamma}(p_1, p_2; q)
-i \int \frac{d^4 p}{(2 \pi)^4} \, \widetilde{\Gamma}(p_1, p; q) \,
\phi_2(p) \, \Gamma(p, p_2; q) \\
+ \frac{z_{\kf}^2 \kf^2}{(2 \pi)^3} \int d\Omega_{\mathbf{p}} \,
\widetilde{\Gamma}(p_1, p; q) \, \frac{|\mathbf{q}| 
\cos\theta_{\mathbf{p}\mathbf{q}}}{
\omega - \vf |\mathbf{q}| \cos\theta_{\mathbf{p}\mathbf{q}}}
\, \Gamma(p, p_2; q) \,.
\label{BSZS2}
\end{multline}
In the limit $|\mathbf{q}|/\omega \to 0$, we have for the
quasiparticle-quasihole irreducible part of the four-point function
$\Gamma^\omega$,
\begin{equation}
\Gamma^\omega(p_1,p_2) = \widetilde{\Gamma}(p_1, p_2)
-i \int \frac{d^4 p}{(2 \pi)^4} \, \widetilde{\Gamma}(p_1, p) \,
\phi_2(p) \, \Gamma^\omega(p, p_2) \,.
\label{eq:Gamma_w}
\end{equation}
Using $\Gamma^\omega$, we can then eliminate the ZS-channel irreducible
four-point function $\widetilde{\Gamma}$ and the background term $\phi_2$
to write the Bethe-Salpeter equation in the form
\begin{equation}
\Gamma(p_1,p_2; q) = \Gamma^\omega(p_1,p_2)
+ \frac{z_{\kf}^2 \kf^2}{(2 \pi)^3} \int d\Omega_{\mathbf{p}}
\Gamma^\omega(p_1, p) \frac{|\mathbf{q}| 
\cos\theta_{\mathbf{p}\mathbf{q}}}{
\omega - \vf |\mathbf{q}| \cos\theta_{\mathbf{p}\mathbf{q}}}
\Gamma(p, p_2; q) .
\label{qp-scatt-eq}
\end{equation}
In the limit $\omega/|\mathbf{q}| \to 0$ one then finds
\begin{equation}
\Gamma^q(p_1,p_2) = \Gamma^\omega(p_1,p_2)
- \frac{z_{\kf}^2 m^* \kf}{(2 \pi)^3} \int d\Omega_{\mathbf{p}} \,
\Gamma^\omega(p_1, p) \, \Gamma^q(p, p_2) \,,
\label{eq:BS_Gamma_k} 
\end{equation}
which describes the scattering of quasiparticles. By identifying the 
quasiparticle interaction with (as justified in Section~\ref{functional})
\begin{equation}
f_{\mathbf{p}_1 \mathbf{p}_2} = z_{\kf}^2 \, \Gamma^{\omega}(p_1, p_2)
\bigr|_{\omega_1=\varepsilon_{\mathbf{p}_1}, \, \omega_2=\varepsilon_{\mathbf{p}_2}} \,,
\label{eq:qp_identifications}
\end{equation}
and the quasiparticle scattering amplitude with
\begin{equation}
a_{\mathbf{p}_1 \mathbf{p}_2} = z_{\kf}^2 \, \Gamma^q(p_1, p_2)
\bigr|_{\omega_1=\varepsilon_{\mathbf{p}_1}, \, \omega_2=\varepsilon_{\mathbf{p}_2}} \,,
\end{equation}
the Bethe-Salpeter equation for the quasiparticle scattering amplitude
reads
\begin{equation}
a_{\mathbf{p}_1 \sigma_1 \mathbf{p}_2 \sigma_2} = 
f_{\mathbf{p}_1 \sigma_1 \mathbf{p}_2 \sigma_2} - \frac{N(0)}{8 \pi} \, 
\sum_\sigma
\int d\Omega_{\mathbf{p}} \, a_{\mathbf{p}_1 \sigma_1 \mathbf{p} \sigma} \, 
f_{\mathbf{p} \sigma \mathbf{p}_2 \sigma_2} \,,
\label{landau-eq}
\end{equation}
where we have reintroduced spin indices. By expanding the angular
dependence of the quasiparticle scattering amplitude on Legendre 
polynomials,
\begin{equation}
a_{\mathbf{p}_1 \mathbf{p}_2}^{s/a} = a^{s/a}(\cos \theta_{\mathbf{p}_1 \mathbf{p}_2}) 
= \sum_l \, a_l^{s/a} \, P_l(\cos \theta_{\mathbf{p}_1 \mathbf{p}_2}) \,,
\label{eq:a_expansion}
\end{equation}
and using the corresponding expansion of the quasiparticle interaction,
Eq. (\ref{eq:f_expansion}), the quasiparticle scattering equation, 
Eq.~(\ref{landau-eq}), can be solved analytically for each value of
$l$,
\begin{equation}
A_l^{s,a} = \cfrac{F_l^{s,a}}{1+\cfrac{F_l^{s,a}}{2 l+1}} \,,
\label{qpsol}
\end{equation}
with $A_l^{s,a} = N(0) \, a_l^{s,a}$. This analytic solution of the 
quasiparticle scattering equation, Eq.~(\ref{qp-scatt-eq}), is in 
general possible only in the limit $q \to 0$.

Quasiparticles are fermionic excitations and therefore obey the Pauli 
principle. This imposes nontrivial constraints on the Landau parameters,
as can be seen by the following argument. The full four-point function
must be antisymmetric under exchange of two particles in the initial
or final states,
\be
\parbox{3.1cm}{\includegraphics[scale=0.82,clip=]{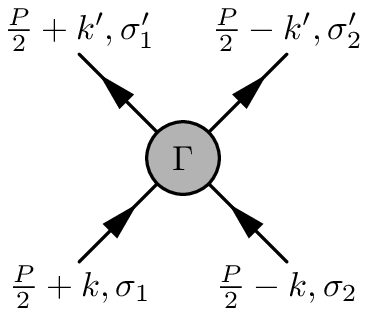}}
= \hspace*{2mm} - \hspace*{4mm}
\parbox{3.1cm}{\includegraphics[scale=0.82,clip=]{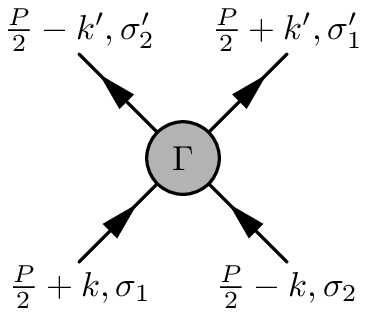}} \,,
\ee
and therefore the forward scattering amplitude, $q=0$, of identical 
particles, $q'=0$ and $\sigma_1=\sigma_2$, must vanish. This implies
$a_{\mathbf{p}_1 \sigma_1 \mathbf{p}_1 \sigma_1} =0$, which leads to the 
Pauli-principle sum rule~\cite{Landau3,AK}
\begin{equation}
\sum_l \, \bigl( A_l^s + A_l^a \bigr) = 0 \,.
\label{sumrule}
\end{equation}
The relations given in this section can be generalized to include
isospin and tensor forces~\cite{GerryPR,tensor,BSJ,FD}, which play an
important role in nuclear systems.

\subsection{Functional approach}
\label{functional}

Functional methods provide a powerful tool for studying many-body
systems. We start by discussing the two-particle irreducible (2PI)
effective action. This provides a useful framework for an RG approach
to many-body systems. For simplicity, we first consider bosonic
systems described by a scalar field $\phi$ and generalize the results
later to fermions. Our discussion follows Ref.~\cite{CJT}. We start
with the expression for the generating functional $W$ for connected
$N$-point functions
\begin{multline}
W[J,K] = - \ln \int \mathcal{D} \phi \, \exp\biggl[
i S[\phi] - \int d^4 x \, \phi(x) \, J(x) \\
- \frac{1}{2} \int d^4 x \, d^4 y \, \phi(x) \, K(x,y) \, \phi(y)
\biggr] \,,
\end{multline}
where $\int \mathcal{D} \phi$ denotes a functional integral over the
field $\phi$, the action is given by $S[\phi] = \int d^4 x \,
\mathcal{L}(\phi(x))$, and $J(x)$ and $K(x,y)$ are external sources.
In thermodynamic equilibrium, the space-time integral is
\be
\int d^4 x = \int_0^{-i \beta} dt \int d^3 x \,,
\ee
with inverse temperature $\beta = 1/T$. By taking functional
derivatives with respect to $J$ and $K$, we obtain the
expectation value of the field $\overline{\phi} = \langle \phi
\rangle$ and the Green's function $G$ respectively,
\begin{align}
\frac{\delta W[J,K]}{\delta J(x)} &= \langle \phi(x) \rangle =
\overline{\phi}(x) \,, \\[1mm]
\frac{\delta W[J,K]}{\delta K(x,y)} &= \frac{1}{2} \Bigl( 
\overline{\phi}(x) \overline{\phi}(y) + G(x,y) \Bigr) \,.
\end{align}
A double Legendre transform leads to the 2PI effective action 
$\Gamma[\overline{\phi},G]$,
\begin{multline}
\Gamma[\overline{\phi},G] = W[J,K] - \int d^4 x \, \overline{\phi}(x)
\, J(x) \\
- \frac{1}{2} \int d^4 x \ d^4 y \, \Bigl[ \overline{\phi}(x) \,
\overline{\phi}(y) + G(x,y) \Bigr] K(x,y) \,,
\end{multline}
which is stationary with respect to variations of
$\overline{\phi}$ and $G$ for vanishing sources,
\begin{align}
\frac{\delta \Gamma[\overline{\phi},G]}{\delta \overline{\phi}(x)} &= 
- J(x) - \int d^4 y \, K(x,y) \, \overline{\phi}(y) \overset{J=K=0}{=} 0 \,,
\label{gap} \\[1mm]
\frac{\delta \Gamma[\overline{\phi},G]}{\delta G(x,y)} &= -
\frac{1}{2} \, K(x,y) \overset{J=K=0}{=} 0 \,.
\label{Dyson}
\end{align}

An explicit form for $\Gamma$ in terms of the expectation value
$\overline{\phi}$ and the exact Green's function $G$ can be
constructed following Refs.~\cite{CJT,LW,Baym}. This leads to
\be
\Gamma[\overline{\phi},G] = -i S[\overline{\phi}] + 
{\rm Tr} \ln G^{-1} + {\rm Tr} \Bigr[ (G_0^{-1}
- G^{-1}) G \Bigr] - \Phi[\overline{\phi},G] \,.
\label{LWfunct}
\ee
Here the functional $\Phi$ is the sum of all 2PI skeleton diagrams
(for a discussion of diagrams, see below) and the trace is a short-hand
notation for
\be
{\rm Tr} =  \int d^4 x \: {\rm tr} = \int_0^{-i \beta} dt \int d^3 x
\: {\rm tr} \,,
\label{trace}
\ee
where ${\rm tr}$ denotes the trace over the internal degrees of
freedom, such as spin and isospin. The stationarity of the 2PI
effective action in the absence of sources, Eqs.~(\ref{gap}) 
and~(\ref{Dyson}), leads to the gap and Dyson equations, respectively.
With the explicit form for $\Gamma$,
Eq.~(\ref{LWfunct}), the Dyson equation is given by
\be
\frac{\delta \Gamma[\overline{\phi},G]}{\delta G} = 
- G^{-1} + G_0^{-1} + \frac{\delta \Phi[\overline{\phi},G]}{\delta G}
= 0 \,,
\ee
which implies
\be
G^{-1} = G_0^{-1} - \Sigma \,,
\label{eq:self-energy1}
\ee
with the self-consistently determined self-energy $\Sigma$ defined by
\be
\Sigma = \frac{\delta \Phi[\overline{\phi},G]}{\delta G} \,.
\label{eq:self-energy2}
\ee

At the stationary point, the 2PI effective action $\Gamma$ is
proportional to the thermodynamic potential $\Omega = T 
\Gamma$ (in units with volume $V=1$)~\cite{Wetterich}:
\be
\Omega(\mu,T) =  -\int d^3 x \, \mathcal{L}(\overline{\phi}(x))
+ \Omega_0(\mu,T) + T \, \Bigl[ {\rm Tr} \ln ( 1 - G_0 \Sigma ) + 
{\rm Tr} \Sigma G - \Phi(\overline{\phi},G) \Bigr] \,,
\label{th-pot-bos}
\ee
where we have introduced the thermodynamic potential of the
non-interacting Bose gas,
\be
\Omega_0(\mu,T) = T \, {\rm Tr} \ln G_0^{-1} = T \, {\rm tr}
\int \frac{d{\bf p}}{(2 \pi)^3} \, \ln \biggl[ 1 - \exp \Bigl[
- \beta \, \Bigl( \frac{{\bf p}^2}{2m} - \mu \Bigr) \Bigr] \biggr]
\,,
\ee
and the third term on the right-hand side of Eq.~(\ref{th-pot-bos})
can also be written as $- T \, {\rm Tr} \ln G_0^{-1} + T \, {\rm Tr} \ln
G^{-1}$, so that $\Omega_0(\mu,T) + T \, {\rm Tr} \ln ( 1 - G_0 \Sigma ) =
T \, {\rm Tr} \ln G^{-1}$. Finally, one can verify that the form of the 2PI
effective action given by Eq.~(\ref{th-pot-bos}), where the
self-energy enters explicitly, is stationary with respect to
independent variations of $G$ and $\Sigma$.

\begin{figure}[t]
\begin{center}
\includegraphics[scale=0.85,clip=]{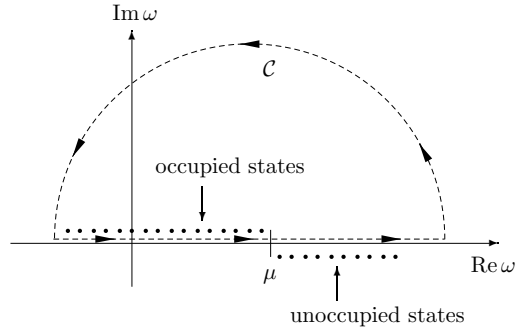}
\end{center}
\caption{Integration contour in the complex $\omega$ plane at $T=0$.
\label{fig:contour}}
\end{figure}

Similarly one has for the thermodynamic potential of a system
consisting of fermions (where the expectation value $\langle \psi
\rangle$ vanishes in the absence of sources)
\be
\Omega (\mu,T)= - T \, \Bigl[ {\rm Tr} \ln G^{-1} + {\rm Tr} \Sigma G 
- \Phi[G] \Bigr] \,.
\label{E_fermions}
\ee
Moreover the energy of a fermionic system can be expressed at $T=0$ in a
form similar to Eq.~(\ref{th-pot-bos}),
\be
E = E_0 - {\rm Tr} \ln (1 - G_0 \Sigma) - {\rm Tr} \Sigma G + \Phi[G] \,.
\label{ener-zero-temp}
\ee
Here $E_0 = {\rm tr} \int \frac{d{\bf p}}{(2 \pi)^3} \, \frac{{\bf 
p}^2}{2m} \, n^0_{{\bf p} \sigma}$ is the energy of the non-interacting
system, and at $T=0$ the trace in Eq.~(\ref{ener-zero-temp}) is given by
\be
{\rm Tr} = {\rm tr} \int \frac{d^4 p}{(2 \pi)^4 i}
= {\rm tr} \oint_{\mathcal{C}}
\frac{d \omega}{2 \pi i} \, \frac{d{\bf p}}{(2 \pi)^3} \,,
\label{eq:zero-T-trace}
\ee
where the integration contour $\mathcal{C}$ is shown in
Fig.~\ref{fig:contour}.

\begin{figure}[t]
\begin{center}
\includegraphics[scale=0.85,clip=]{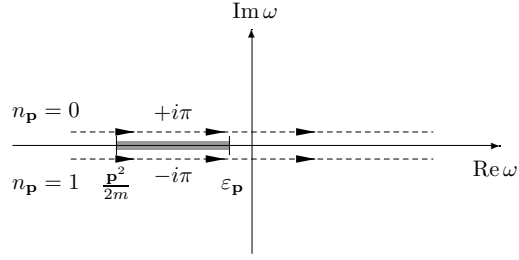}
\end{center}
\caption{Analytic structure of $\ln(G_0 \, G^{-1})$ in the complex
$\omega$ plane for $\varepsilon_{\mathbf{p}}>\frac{\mathbf{p}^2}{2m}$.
The upper (lower) contour is for $n_{\bf p}=0$ ($n_{\bf p}=1$) and
the shaded region represents the complex cut on the real axis.
\label{fig:ana_GG0}}
\end{figure}

Next, we use the functional integral approach to relate the quasiparticle
interaction to the quasiparticle-quasihole irreducible part of the
four-point function $\Gamma^\omega$ following Ref.~\cite{Nozieres}.
In the quasiparticle approximation, the full propagator takes the
form of Eq.~(\ref{eq:G_qp}),
\be
G(\omega,\mathbf{p}) = \frac{1}{G_0^{-1}(\omega,\mathbf{p}) -
\Sigma(\omega,\mathbf{p})} = \frac{z_{\mathbf{p}}}{\omega - 
\varepsilon_{\mathbf{p}} + i \delta_{\mathbf{p}}} 
+ \phi(\omega,\mathbf{p}) \,,
\label{qpprop}
\ee
where we have neglected the imaginary part of the self-energy for
excitations close to the Fermi surface. As shown in
Section~\ref{microscopic}, the quasiparticle
energy $\varepsilon_{\mathbf{p}}$ is given by the self-consistent
solution to the Dyson equation, Eq.~(\ref{eq:dyson-qp-energy}),
\be
\varepsilon_{\mathbf{p}} = \frac{\mathbf{p}^2}{2m} +
\Sigma(\varepsilon_{\mathbf{p}},\mathbf{p}) \,,
\label{qpenergies}
\ee
and the single-particle strength $z_{\mathbf{p}}$ by Eq.~(\ref{zfact}).
When a quasiparticle with momentum $\mathbf{p}$ is added to the
system, the state is changed from unoccupied to occupied, so that
$\delta_{\mathbf{p}} = \delta \rightarrow - \delta$, with positive
infinitesimal $\delta > 0$.  Because the 2PI effective action is
stationary with respect to independent variations of $G$ and $\Sigma$,
we only need to consider changes of $E_0$ and those induced by
variations of $G_0$ in Eq.~(\ref{ener-zero-temp}). In the
quasiparticle approximation, we have for the argument of the
logarithm,
\be
1-G_0(p) \Sigma(p) = G_0(p) \, G^{-1}(p) = \frac{1}{z_{\mathbf{p}}} \,
\frac{\omega - \varepsilon_{\mathbf{p}} + i \delta_{\mathbf{p}}}{
\omega - \frac{\mathbf{p}^2}{2m} + i \delta_{\mathbf{p}}} + 
\text{smooth parts.}
\ee
Consider the case $\varepsilon_{\mathbf{p}} > 
\frac{\mathbf{p}^2}{2m}$. Then the real part of $G_0 \, G^{-1}$
is negative for $\frac{\mathbf{p}^2}{2m} < \omega < 
\varepsilon_{\mathbf{p}}$, resulting in a cut on the real energy,
as shown in Fig. \ref{fig:ana_GG0}. When a particle is added to
the system, the integration contour changes from above the cut
to below, and $\ln(G_0 \, G^{-1})$ changes by $-2 \pi i$ for
$\frac{\mathbf{p}^2}{2m} < \omega < \varepsilon_{\mathbf{p}}$.
As a result, the change in the energy of the system is given by
\begin{align}
\frac{\delta E}{\delta n_{\mathbf{p}}} &= 
\frac{\delta E_0}{\delta n_{\mathbf{p}}}
- \frac{\delta}{\delta n_{\mathbf{p}}} \biggl( \,
\oint_{\mathcal{C}} \frac{d\omega}{2 \pi i} \, \frac{d{\bf p}}{(2 \pi)^3} \,
\ln \bigl( G_0(p) \, G^{-1}(p) \bigr) \biggr) \,, \\[1mm]
&=\frac{\mathbf{p}^2}{2m} + \biggl( \varepsilon_{\mathbf{p}} 
- \frac{\mathbf{p}^2}{2m} \, \biggr) = \varepsilon_{\mathbf{p}} \,.
\end{align}
This variation is the quasiparticle energy, as postulated by Landau.

\begin{figure}[t]
\begin{center}
\parbox{3.0cm}{\includegraphics[scale=0.9]{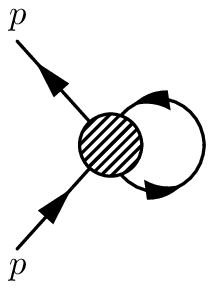}}
\hspace*{6mm}
\parbox{3.0cm}{\includegraphics[scale=0.9]{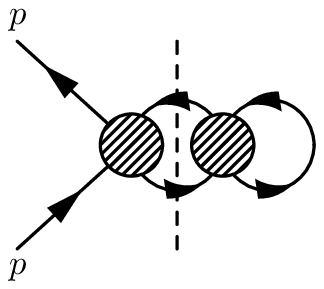}}
\end{center}
\caption{Only 2PI skeleton diagrams contribute to the self-energy
$\Sigma$, as shown on the left. Non-skeleton diagrams, such as the
one on the right, are not part of $\Sigma$. Therefore, the kernel
$\widetilde{\Gamma} = \delta \Sigma/\delta G$ (denoted by the 
shaded blob) is particle-hole irreducible in the ZS channel.
\label{Sigmairred}}
\end{figure}

As discussed in Section~\ref{basic}, the quasiparticle interaction
$f_{\mathbf{p}_1 \mathbf{p}_2}$ is obtained by an additional variation
with respect to the occupation number $n_{\mathbf{p}_2}$,
\be
f_{\mathbf{p}_1 \mathbf{p}_2} = \frac{\delta^2 E}{\delta n_{\mathbf{p}_1}
\, \delta n_{\mathbf{p}_2}} = \frac{\delta \varepsilon_{\mathbf{p}_1}}{
\delta n_{\mathbf{p}_2}} \,.
\ee
Using the Dyson equation, Eq.~(\ref{qpenergies}), the variation of the
quasiparticle energy $\varepsilon_{\mathbf{p}_1}$ with respect to the
occupation number $n_{\mathbf{p}_2}$ yields
\be
\frac{\delta \varepsilon_{\mathbf{p}_1}}{\delta n_{\mathbf{p}_2}}
= \frac{\delta \Sigma(\omega_1,\mathbf{p}_1)}{
\delta n_{\mathbf{p}_2}} \biggr|_{\omega_1 = \varepsilon_{\mathbf{p}_1}}
+ \frac{\partial \Sigma(\omega_1,\mathbf{p}_1)}{
\partial \omega_1} \biggr|_{\omega_1 = \varepsilon_{\mathbf{p}_1}}
\frac{\delta \varepsilon_{\mathbf{p}_1}}{\delta n_{\mathbf{p}_2}} \,.
\ee
This can be expressed with the single-particle strength
$z_{\mathbf{p}}$ as
\be
\frac{\delta \varepsilon_{\mathbf{p}_1}}{\delta n_{\mathbf{p}_2}}
= z_{\mathbf{p}_1} \, \frac{\delta \Sigma(\omega_1,\mathbf{p}_1)}{
\delta n_{\mathbf{p}_2}} \biggr|_{\omega_1 = \varepsilon_{\mathbf{p}_1}}
\,. \label{eq:var_epsilon}
\ee
Furthermore, it follows from the definition of $\Sigma$ through
the $\Phi$ functional, $\Sigma(p) = \delta \Phi[G]/\delta G(p)$,
that the self-energy consists of skeleton diagrams. Therefore,
we can write
\be
\frac{\delta \Sigma(\omega_1,\mathbf{p}_1)}{
\delta n_{\mathbf{p}_2}} \biggr|_{\omega_1 = \varepsilon_{\mathbf{p}_1}} =
\int \frac{d^4 p}{(2 \pi)^4 i} \:
\frac{\delta \Sigma(\omega_1,\mathbf{p}_1)}{\delta G(p)}
\, \frac{\delta G(p)}{\delta n_{\mathbf{p}_2}} 
\biggr|_{\omega_1 = \varepsilon_{\mathbf{p}_1}} \,.
\label{eq:var_sigma}
\ee
The variation of $\Sigma$ with respect to $G$ selects one of the
internal lines of the diagrams contributing to the self-energy.
As a result, the kernel
\be
\widetilde{\Gamma}(p_1,p_2) = \frac{\delta \Sigma(p_1)}{\delta G(p_2)} \,,
\ee
must be particle-hole irreducible in the ZS channel.
Otherwise the corresponding diagram in $\Sigma$
would not have been a 2PI skeleton diagram. This is illustrated in
Fig.~\ref{Sigmairred}.

Using the quasiparticle form of the full propagator,
Eq.~(\ref{qpprop}), there are two contributions to $\delta
G(p_1)/\delta n_{\mathbf{p}_2}$ (for details, see also
Ref.~\cite{Nozieres}). One for $p_1=p_2$ ($\omega_2 = 
\varepsilon_{\mathbf{p}_2}$), which results in a shift of
the quasiparticle pole across the integration contour, and one for
$p_1 \neq p_2$, which corresponds to a variation of the self-energy,
\be
\frac{\delta G(p_1)}{\delta n_{\mathbf{p}_2}} = (2 \pi)^4 i \,
z_{\mathbf{p}_2} \, \delta(\mathbf{p}_1 - \mathbf{p}_2) 
\delta(\omega_1 - \varepsilon_{\mathbf{p}_2}) + 
\frac{\delta \Sigma (p_1)}{\delta n_{\mathbf{p}_2}}
\, G^2(p_1) \,.
\label{eq:delta_G}
\ee
The $G^2(p_1)$ part in the second term is equivalent to the
non-singular contribution $\phi_2(p_1)$ of Eq.~(\ref{eq:GG}).
By inserting this expression for $\delta G(p_1)/\delta
n_{\mathbf{p}_2}$ in Eq.~(\ref{eq:var_sigma}), one finds the
integral equation
\begin{align}
Y(p_1,p_2) &= \frac{1}{z_{\mathbf{p}_2}} \frac{\delta \Sigma(\omega_1,
\mathbf{p}_1)}{\delta n_{\mathbf{p}_2}} \biggr|_{\omega_1 =
\varepsilon_{\mathbf{p}_1}} \, \\[1mm]
&= \widetilde{\Gamma}(p_1,p_2) +
\int \frac{d^4 p}{(2 \pi)^4 i} \:
\widetilde{\Gamma}(p_1,p) \,  G^2(p) \, Y(p,p_2) \,.
\end{align}
A comparison with Eq.~(\ref{eq:Gamma_w}) leads to the 
identification
\be
Y(p_1,p_2) = \Gamma^{\omega}(p_1,p_2) = 
\frac{1}{z_{\mathbf{p}_2}} \frac{\delta \Sigma(\omega_1,
\mathbf{p}_1)}{\delta n_{\mathbf{p}_2}} \biggr|_{\omega_1 =
\varepsilon_{\mathbf{p}_1}} \,.
\label{eq:Y-amplitude}
\ee
Using Eq.~(\ref{eq:var_epsilon}), this implies
\be
f_{\mathbf{p}_1 \mathbf{p}_2} = \frac{\delta
\varepsilon_{\mathbf{p}_1}}{\delta n_{\mathbf{p}_2}} = 
z_{\mathbf{p_1}} z_{\mathbf{p}_2} \, 
\Gamma^{\omega}(p_1,p_2) =
\parbox{2.7cm}{\includegraphics[scale=0.82,clip=]{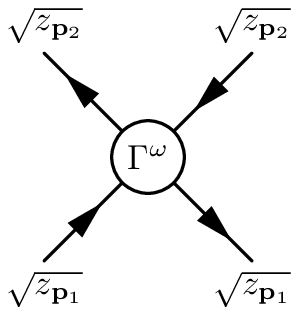}} \,.
\label{eq:qp-int}
\end{equation}
This provides a microscopic basis for calculating the quasiparticle energy
$\varepsilon_{\mathbf{p}} = \frac{\delta E}{\delta n_{\mathbf{p}_1}}$
and the quasiparticle interaction $f_{\mathbf{p}_1 \mathbf{p}_2} = 
\frac{\delta^2 E}{\delta n_{\mathbf{p}_1} \delta n_{\mathbf{p}_2}}$
and a justification for the identification of the quasiparticle
interaction as in Eq.~(\ref{eq:qp_identifications}).

\begin{figure}[t]
\begin{center}
\begin{tabular}{c|c|c}
$E^{(2)}$ & $\varepsilon^{(2)}_{\mathbf{p}}$ & $f^{(2)}_{\mathbf{p}_1
\mathbf{p}_2}$ \\[1mm] \hline
\parbox{2.8cm}{\vspace*{0.5mm}
\centering \includegraphics[scale=0.76]{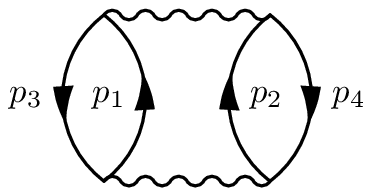}} &
\: \parbox{2.2cm}{\vspace*{1mm}
\centering \includegraphics[scale=0.80]{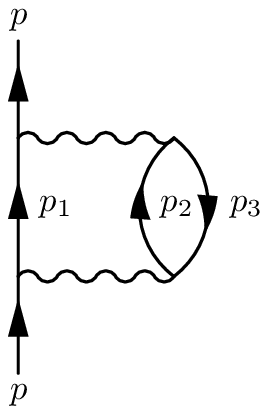}} &
\: \parbox{2.6cm}{
\centering \includegraphics[scale=0.84]{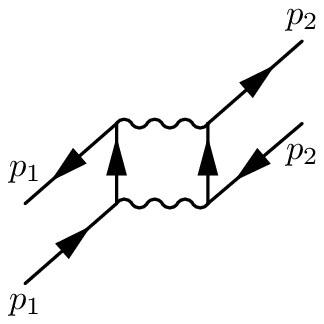}}
\: \parbox{2.8cm}{
\centering \includegraphics[scale=0.84]{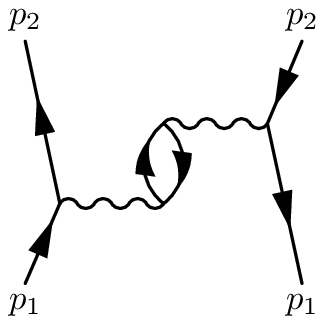}}
\end{tabular}
\end{center}
\caption{The variation of the second-order energy diagram $E^{(2)}$
(left) with respect to the quasiparticle distribution function
yields the second-order contribution to the quasiparticle energy
$\varepsilon^{(2)}_{\mathbf{p}}$ given by the self-energy diagram
(middle) and the corresponding two-hole--one-particle diagram. The
second variation gives the second-order contributions to the
quasiparticle interaction (right). These include the
particle-particle and particle-hole diagrams shown and a
particle-hole diagram that is obtained from the particle-particle
one by reversing the arrow on the $p_2$ line (plus the diagrams
obtained from the two-hole--one-particle self-energy
contribution).\label{fig:variation}}
\end{figure}

\begin{figure}[t]
\begin{center}
\parbox{2.8cm}{\includegraphics[scale=0.84]{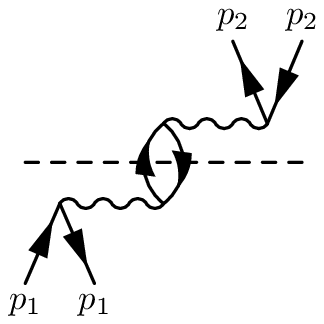}}
\hspace*{8mm}
\parbox{2.6cm}{\includegraphics[scale=0.84]{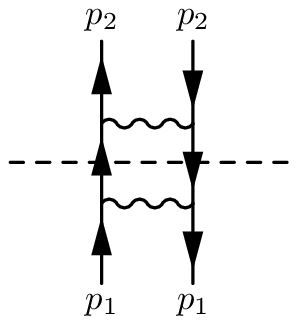}}
\end{center}
\caption{Second-order diagrams that are not generated by variations
of the energy diagram.\label{fig:reducible}}
\end{figure}

The contributions to the quasiparticle interaction can be understood
by considering the variation of the second-order energy diagram, as
shown in Fig.~\ref{fig:variation}. The resulting diagrams in
$f_{\mathbf{p}_1 \mathbf{p}_2}$ are quasiparticle-quasihole reducible
in the BCS and ZS$'$ channels, but irreducible in the ZS channel (see
Fig.~\ref{channels} for a definition of the BCS, ZS and ZS$'$
channels). The ZS reducible diagrams, which are shown for the
second-order example in Fig.~\ref{fig:reducible}, are included in the
scattering amplitude $a_{\mathbf{p}_1 \mathbf{p}_2}$ but not in the
quasiparticle interaction. Because the ZS and ZS$'$ channels are
related by exchange~\cite{BB}, the quasiparticle scattering amplitude
is antisymmetric, but the quasiparticle interaction is not.

\begin{figure}[t]
\begin{center}
\parbox{2.1cm}{\includegraphics[scale=0.82,clip=]{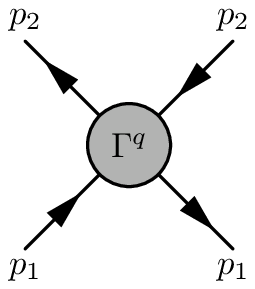}}
$\quad = \quad$
\parbox{2.1cm}{\includegraphics[scale=0.82,clip=]{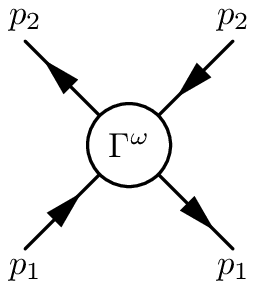}}
$\quad + \quad$
\parbox{2.1cm}{\includegraphics[scale=0.82,clip=]{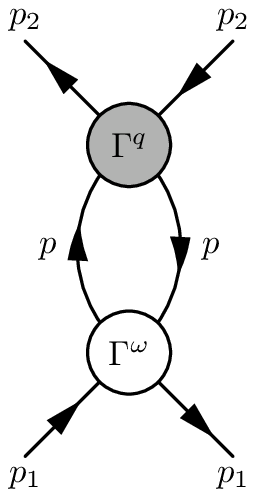}}
\end{center}
\caption{The Bethe-Salpeter equation for the quasiparticle scattering
amplitude, Eq.~(\ref{eq:BS_Gamma_k}), which sums diagrams reducible
in the ZS channel. The intermediate-state propagators include only
the quasiparticle-quasihole part. All other contributions are
included in $\Gamma^\omega$.\label{fig:BS}}
\end{figure}

\begin{figure}[t]
\begin{center}
\parbox{2.1cm}{\includegraphics[scale=0.82,clip=]{Gammak1.eps}}
$\quad = \quad$
\parbox{2.1cm}{\includegraphics[scale=0.82,clip=]{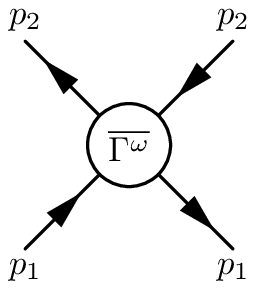}}
$\quad + \quad$
\parbox{3.8cm}{\includegraphics[scale=0.82,clip=]{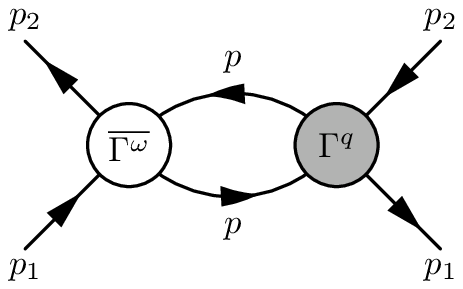}}
\end{center}
\caption{The Bethe-Salpeter equation for the quasiparticle scattering
amplitude in the exchange (ZS$'$) channel. In this channel, the solution
requires as input the quasiparticle scattering amplitude $\Gamma^q$ and
the quasiparticle interaction $\Gamma^\omega$ at finite $q$. As in 
Fig.~\ref{fig:BS}, the intermediate-state propagators include only
the quasiparticle-quasihole part.\label{fig:BS2}}
\end{figure}

The quasiparticle-quasihole reducible diagrams in the ZS channel are
summed by the Bethe-Salpeter equation, Eq.~(\ref{eq:BS_Gamma_k}),
which yields the fully reducible four-point function $\Gamma^q$, given
the quasiparticle-quasihole irreducible one $\Gamma^\omega$, as shown
in Fig.~\ref{fig:BS}. The fully reducible four-point function
$\Gamma^q$ corresponds to the quasiparticle scattering amplitude. The
four-point function can also be obtained by summing diagrams that are
quasiparticle-quasihole reducible in the ZS$'$ channel. The
corresponding Bethe-Salpeter equation is shown in Fig.~\ref{fig:BS2},
where the irreducible term is the quasiparticle interaction in the
ZS$'$ channel, the exchange of $\Gamma^\omega$ denoted by
$\overline{\Gamma^\omega}$.

\begin{figure}[t]
\begin{center}
\parbox{2.1cm}{\includegraphics[scale=0.82,clip=]{Gammak2.eps}}
$\quad = \quad$
\parbox{2.1cm}{\includegraphics[scale=0.82,clip=]{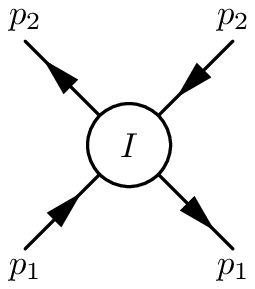}}
$\quad + \quad$
\parbox{3.8cm}{\includegraphics[scale=0.82,clip=]{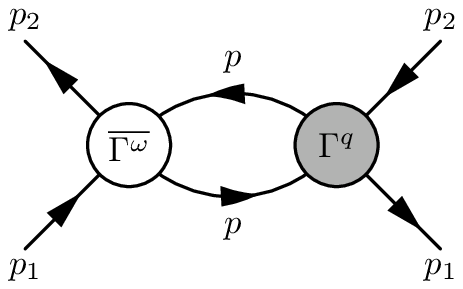}}
\end{center}
\caption{Integral equation for the quasiparticle interaction
$\Gamma^\omega$ that sums quasiparticle-quasihole reducible diagrams
in the ZS$'$ channel. The second term on the right hand side is
the induced interaction of Babu and Brown~\cite{BB}.\label{fig:induced}}
\end{figure}

The kinematics in the integral term on the right-hand side of
Fig.~\ref{fig:BS2} requires as input the quasiparticle scattering
amplitude $\Gamma^q$ and the quasiparticle interaction $\Gamma^\omega$
at finite $q$. Therefore, we can generalize $\Gamma^q$ on the
left-hand side to finite $q$. If we then take the limit
$|\mathbf{q}|/\omega \to 0$, all quasiparticle-quasihole reducible
terms in the ZS channel vanish. In this limit, $\Gamma^q$ on the 
left-hand side of Fig.~\ref{fig:BS2} is replaced by $\Gamma^\omega$, and
the first term on the right hand side, $\overline{\Gamma^\omega}$, is
reduced to the driving term $I$, which is quasiparticle-quasihole
irreducible in both ZS and ZS$'$ channels.  As a result, we obtain an
integral equation for $\Gamma^\omega$ that sums
quasiparticle-quasihole reducible diagrams in the ZS$'$ channel. This
is shown diagrammatically in Fig.~\ref{fig:induced}, where the second
term on the right hand side is the induced interaction of Babu and
Brown~\cite{BB} (see also Ref.~\cite{indint}). For $p_1 = p_2$, this
integral equation has the simple form, \be \Gamma^\omega(p_1,p_1) =
I(p_1,p_1) - P_{\bm \sigma} \, \frac{z_{\kf}^2 m^* \kf}{(2 \pi)^3} \,
\sum\limits_\sigma \int d\Omega_{\mathbf{p}} \, \Gamma^\omega(p_1, p)
\, \Gamma^q(p, p_1) \,.  \ee The induced interaction accounts for the
contributions to the quasiparticle interaction due to the polarization
of the medium and is necessary for the antisymmetry of the
quasiparticle scattering amplitude $\Gamma^q$.

\section{Functional RG approach to Fermi liquid theory}
\label{RGnm}

In this section we apply the functional RG~\cite{Wetterich93} to
calculate the properties of a Fermi liquid. The basic idea is to
renormalize the quasiparticle energy and the quasiparticle
interaction, as one sequentially integrates out the excitations of the
system. Thereby, one starts with the high-lying states and integrates
down to low excitation energy. The functional RG leads to an infinite
set of coupled differential equations for the $n$-point functions,
which in practical calculations must be truncated. We return to this
question below, when we discuss applications.

\begin{figure}[t]
\begin{center}
\input{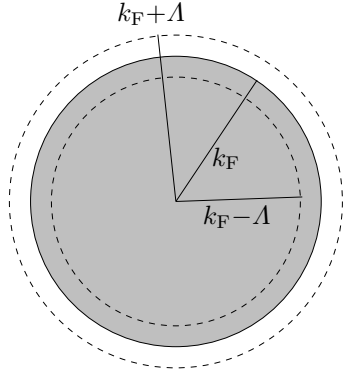}
\end{center}
\caption{The cutoffs above and below the Fermi surface separate high- 
and low-lying excitations of the many-fermion system.\label{sea}}
\end{figure}

In zero-temperature Fermi systems, the low-lying states are those near
the Fermi surface. Consequently, at some intermediate step of the
calculation, the high-lying states, far above and far below the Fermi
surface have been integrated out, while the states near the Fermi
surface are not yet included. This is schematically illustrated in
Fig.~\ref{sea}. However, in the first part of this section we keep the
discussion more general and allow for finite temperatures. In the
second part, we apply the RG approach to a Fermi liquid at zero
temperature and also specify the detailed form of the regulator
adapted to Fermi systems at zero temperature.

The Green's function 
\begin{equation}
G_{\Lambda 0}^{-1}(\omega,\mathbf{p}) = \omega -
\frac{\mathbf{p}^2}{2m} - R_{\Lambda}(\mathbf{p}) \,,
\label{eq:prop-reg}
\end{equation}
with regulator $R_{\Lambda}(\mathbf{p})$ defines the free theory at a
finite cutoff scale $\Lambda$. The corresponding effective action is
given by (see Eq.~(\ref{LWfunct}) and the corresponding expression
for fermions,  Eq.~(\ref{E_fermions}))
\begin{equation}
\Gamma_{\Lambda}(\psi^{\ast},\psi,G) = -i
S_{\Lambda}(\psi^{\ast},\psi) - \text{Tr}\ln (-G^{-1}) 
- \text{Tr}[G_{\Lambda 0}^{-1} G  - 1] + \Phi(\psi^{\ast},\psi, G) \,,
\end{equation}
with the action
\begin{equation}
S_{\Lambda}(\psi^{\ast},\psi) = -i \, T \sum_n \int \frac{d^3p}{(2 \pi)^3}
\, \psi^*(\omega_n,\mathbf{p}) \biggl(\omega_n-\frac{\mathbf{p}^2}{2m}
- R_{\Lambda}(\mathbf{p})\biggr) \psi(\omega_n,\mathbf{p}) \,,
\end{equation}
and Matsubara frequency $\omega_{n}=(2n+1) i \pi T$. At the stationary
point, the propagator $G$ satisfies the Dyson equation (see
Eqs.~(\ref{eq:self-energy1}) and~(\ref{eq:self-energy2}))
\begin{equation}
G_{\Lambda}^{-1} = G_{\Lambda 0}^{-1} - \frac{\delta 
\Phi(\psi^{\ast},\psi,G)}{\delta G} \biggr|_{G = G_{\Lambda}} \,,
\label{eq:dyson}
\end{equation}
and therefore acquires a dependence on $\Lambda$.  Due to the
stationarity of the 2PI effective action with respect to variations of
$\psi$, $\psi^{\ast}$, and $G$, only the explicit $\Lambda$ dependence
of the bare Green's function $G_{\Lambda 0}$ contributes to the flow
equation for the effective action:
\begin{eqnarray}
\frac{d \Gamma_{\Lambda}(\psi^{\ast},\psi,G_{\Lambda})}{d \Lambda} 
&=& \text{Tr} \biggl[ \frac{\delta 
\Gamma_{\Lambda}(\psi^{\ast},\psi, G_{\Lambda})}{\delta G^{-1}_{\Lambda 0}
(p)} \frac{d G^{-1}_{\Lambda 0} (p)}{d \Lambda} \biggr] \nonumber
\\[1mm]
&=& \text{Tr} \biggl[ \psi^*_{p} \,
\frac{dR_\Lambda(\mathbf{p})}{d\Lambda} \, \psi_{p}\biggr] +
\text{Tr} \biggl[ G_{\Lambda}(p) \, 
\frac{d R_{\Lambda}(\mathbf{p})}{d \Lambda} \biggr] \,,
\label{eq:flow}
\end{eqnarray}
where we have introduced the short-hand notations $\psi_p =
\psi(\omega_n,\mathbf{p})$ and ${\rm{Tr}} = T \,\sum_n \int
\frac{d \mathbf{p}}{(2\pi)^3}$. The flow equation follows
almost trivially from the stationarity of the 2PI effective action,
while within a 1PI scheme,\footnote{The 1PI
effective action is obtained by constraining the Green's function in
the 2PI effective action to the solution of the Dyson equation,
Eq.~(\ref{eq:dyson}).} the derivation of the flow equation is
somewhat more involved \cite{Wetterich93,BTW}. The flow equation
for the two-point function in the 1PI scheme is obtained by
varying Eq.~(\ref{eq:flow}) with respect to $\psi$ and $\psi^\ast$.
Using
\begin{equation}
\Gamma^{\Lambda(2)}_{p,p} = 
\frac{\delta^2 \Gamma_\Lambda}{\delta \psi^\ast_p\,\delta\psi_p}
= G_{\Lambda}^{-1}(p) \,,
\end{equation}
and
\begin{equation}
\frac{\delta^2 G_\Lambda (p')}{\delta\psi^\ast_p\,\delta\psi_p}
= -G_\Lambda(p') \, \Gamma^{\Lambda(4)}_{p,p,p',p'} \, G_\Lambda(p')
\quad \text{with} \quad
\Gamma^{\Lambda(4)}_{p,p,p',p'} = \frac{\delta^2 
\Gamma^{\Lambda(2)}_{p,p}}{\delta \psi^\ast_{p'}\,\delta \psi_{p'}} \,,
\end{equation}
one finds
\begin{equation}
\frac{d \Gamma^{\Lambda(2)}_{p,p}}{d\Lambda} =
- \frac{d R_\Lambda(\mathbf{p})}{d\Lambda}
- \text{Tr} \biggl[ \Gamma^{\Lambda(4)}_{p,p,p',p'} \, G_\Lambda(p')
\, \frac{dR_\Lambda(\mathbf{p}')}{d\Lambda} \, G_\Lambda(p') \biggr] \,.
\label{eq:flow-one-pi}
\end{equation}

\begin{figure}[t]
\begin{center}
\parbox{0.5cm}{\includegraphics[scale=1.0]{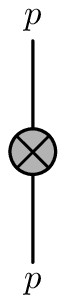}}
\hspace{0.1cm} $-$ \hspace{0.1cm} 
\parbox{0.6cm}{\includegraphics[scale=1.0]{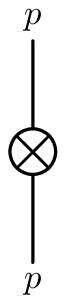}}
\hspace{0.1cm} $=$ \hspace{0.1cm} 
\parbox{2.6cm}{\includegraphics[scale=1.0]{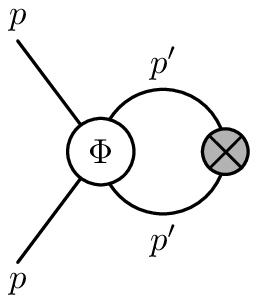}}
\hspace{0.1cm} $=$ \hspace{0.1cm} 
\parbox{2.4cm}{\includegraphics[scale=1.0]{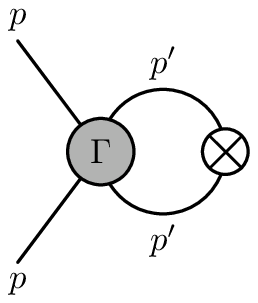}}
\end{center}
\caption{Graphical representation of the 1PI and 2PI flow equations
for the two-point function, Eqs.~(\ref{eq:flow-one-pi})
and~(\ref{eq:flowfromdyson}). The internal lines represent full
propagators $G_\Lambda$. The filled crossed circles represent
$\frac{d \Gamma^{\Lambda (2)}}{d \Lambda}$ and the unfilled ones
$\frac{d G_{\Lambda 0}^{-1}}{d \Lambda} = - \frac{d R_{\Lambda}}{d
\Lambda}$.}
\label{fig:flow_eq}
\end{figure}

Next, we briefly discuss the flow equation in the 2PI scheme and make
a connection between the two schemes for the two-point function. Here
we follow the discussion of Dupuis~\cite{Dupuis}. The starting point
is the observation that the 2PI functional $\Phi$ does not flow, when
$\psi$ and $G$ are treated as free variables
\begin{equation}
\frac{d\Phi(\psi,G)}{d\Lambda} \biggr|_{\psi,G} = 0 \,.
\end{equation}
As discussed in the previous section, the functional $\Phi(\psi,G)$ 
generates the particle-hole irreducible $n$-point functions through
variations with respect to the Green's function (see also 
Ref.~\cite{Baym})
\begin{equation}
\Phi^{\Lambda (2n)}_{p_1,p_2, \ldots, p_{2n-1},p_{2n}} = 
\frac{\delta^{n} \Phi(\psi,G)}{\delta G(p_1, p_2) \cdots \,
\delta G(p_{2n-1}, p_{2n})} \,.
\end{equation}
After the variation, the Green's functions satisfy the Dyson equation,
Eq.~(\ref{eq:dyson}). Consequently, the flow of the 2PI vertices
$\Phi^{(n)}$ results only from the $\Lambda$ dependence of the Green's
function:
\begin{equation}
\frac{d}{d \Lambda} \, \Phi^{\Lambda (2n)}_{p_1,p_2, \ldots, p_{2n-1},p_{2n}} 
= \text{Tr} \biggl[ \Phi^{\Lambda (2n+2)}_{p_1,p_2, \ldots, p_{2n-1},p_{2n},q,q}
\, \frac{d G_{\Lambda}(q)}{d \Lambda} \biggr] \,.
\label{eq:flow_2PI}
\end{equation}
Combined with the self-energy $\Sigma(p) = \Phi^{\Lambda (2)}_{p,p}$,
so that $\Gamma^{\Lambda (2)}_{p,p} = G_{\Lambda 0}^{-1}(p)
-\Phi^{\Lambda (2)}_{p,p}$ and $\frac{d\Gamma^{\Lambda (2)}_{p,p}}{d 
\Lambda} = - \frac{dR_{\Lambda} (\mathbf{p})}{d \Lambda} -
\frac{d \Phi^{\Lambda (2)}_{p,p}}{d \Lambda}$, one finds
\begin{equation}
\frac{d \Gamma^{\Lambda (2)}_{p,p}}{d \Lambda} 
= - \frac{d R_{\Lambda}(\mathbf{p})}{d \Lambda} + \text{Tr}
\biggl[ \Phi^{\Lambda (4)}_{p, p, p', p'} \, G_{\Lambda}(p') \,
\frac{d \Gamma^{\Lambda (2)}_{p',p'}}{d \Lambda} \, G_{\Lambda}(p')
\biggr] \,.
\label{eq:flowfromdyson}
\end{equation}
The flow equations for the two-point function in the 1PI and 2PI
schemes, Eqs.~(\ref{eq:flow-one-pi}) and~(\ref{eq:flowfromdyson}), are
equivalent, as can be shown in a straightforward calculation, making
use of the Bethe-Salpeter equation for scattering of two particles of
vanishing total momentum~\cite{Hebeler},
\begin{equation}
\Gamma^{\Lambda (4)}_{p_1,p_1,p_2,p_2} = \Phi^{\Lambda (4)}_{p_1,p_1,p_2,p_2}
+\text{Tr} \biggl[ \Phi^{\Lambda (4)}_{p_1,p_1,p',p'} \, G_\Lambda(p')
\, G_\Lambda(p') \, \Gamma^{\Lambda (4)}_{p',p',p_2,p_2} \biggr] \,.
\end{equation}
The relation between the two schemes is illustrated diagrammatically
in Fig.~\ref{fig:flow_eq}. The particle-hole reducible diagrams can be
shifted between the four-point function and the regulator insertion on
the fermion line. For more details on the relation between the RG
approaches based on 1PI and 2PI functionals the reader is referred to
Ref.~\cite{Dupuis}.

We are now in a position to connect with Fermi liquid theory, by
deriving a flow equation for the quasiparticle energy closely related
to 1PI equation, Eq.~(\ref{eq:flow-one-pi}). To this end, we first
define the energy functional following Eq.~(\ref{ener-zero-temp})
\begin{equation}
E_\Lambda(G) = E_{\Lambda 0} - \text{Tr} \ln (G_{\Lambda 0} \,
G^{-1} ) - \text{Tr} \bigl[ (G_{\Lambda 0}^{-1}-G^{-1}) \, G \bigr] 
+ \Phi(G) \,,
\label{eq:e-functional}
\end{equation}
which at the stationary point equals the (zero-temperature)
ground-state energy at the cutoff scale $\Lambda$. The trace Tr is
defined as in Eq.~(\ref{eq:zero-T-trace}), $E_{\Lambda 0}$ is the
energy of the non-interacting system at the scale $\Lambda$,
\begin{equation}
E_{\Lambda 0} = \text{tr} \int \frac{d^3p}{(2\pi)^3}
\biggl[ \frac{\mathbf{p}^2}{2m} + R_\Lambda(\mathbf{p}) \biggr]
n_{\mathbf{p},\Lambda}^0 \,,
\label{EL_0}
\end{equation}
with $n_{\mathbf{p},\Lambda}^0 = \theta(\kf - \Lambda - p)$, and
the free Green's function $G_{\Lambda 0}$ is given by
\begin{equation}
G_{\Lambda 0}(\omega,\mathbf{p}) = \frac{1-n_{\mathbf{p},\Lambda}^0}{
\omega-\frac{\mathbf{p}^2}{2m}-R_\Lambda(\mathbf{p})+i\delta}
+ \frac{n_{\mathbf{p},\Lambda}^0}{\omega-\frac{\mathbf{p}^2}{2m}
-R_\Lambda(\mathbf{p})-i\delta} \,. 
\end{equation}
Furthermore, in the quasiparticle approximation the one-particle
Green's function of the interacting system is of the form
\begin{equation}
G_\Lambda(\omega,\mathbf{p}) = z_{\mathbf{p}}
\biggl[\frac{1-n_{\mathbf{p},\Lambda}^0}{\omega-
\tilde{\varepsilon}_{\mathbf{p}}+i\delta}+
\frac{n_{\mathbf{p},\Lambda}^0}{\omega-
\tilde{\varepsilon}_{\mathbf{p}}-i\delta} \biggr]
+ \phi(\mathbf{p},\omega) \,,
\label{G_full}
\end{equation}
where the quasiparticle energy $\tilde{\varepsilon}_{\mathbf{p}}$
is given by
\begin{equation}
\tilde{\varepsilon}_{\mathbf{p}} = \frac{\mathbf{p}^{2}}{2 m}
+ \Sigma_{\Lambda}(\tilde{\varepsilon}_{\mathbf{p}},{\mathbf{p}})
+ R_\Lambda(\mathbf{p}) \,.
\label{dyson-reg}
\end{equation}
The flow equation that follows from the energy functional,
Eq.~(\ref{eq:e-functional}), reads
\begin{equation}
\frac{d E_\Lambda}{d \Lambda} = \frac{d E_{\Lambda 0}}{d \Lambda} 
- \text{Tr} \biggl[ (G_{\Lambda 0} - G_\Lambda) \, \frac{d R_{\Lambda}}{
d \Lambda} \biggr] 
= \text{Tr} \biggl[ G_\Lambda \, \frac{d 
R_{\Lambda}}{d \Lambda} \biggr] \,.
\label{eq:functional}
\end{equation}
By varying Eq.~(\ref{eq:functional}) with respect to the quasiparticle
occupation number, we obtain a flow equation for the quasiparticle energy
\begin{eqnarray}
\frac{d \tilde{\varepsilon}_{\mathbf{p}}}{d \Lambda} &=& 
\frac{\delta}{\delta n_{\mathbf{p}}} \frac{d E_{\Lambda}}{d \Lambda} 
= \frac{\delta}{\delta n_{\mathbf{p}}} \biggl( \text{tr} \int 
\frac{d^4 p'}{(2 \pi)^4 i} \, G_{\Lambda}(p') \, \frac{d R_{\Lambda}
(\mathbf{p}')}{d \Lambda} \biggr) \nonumber \\[1mm]
&=& z_{\mathbf{p}} \, \frac{d R_{\Lambda} (\mathbf{p})}{d \Lambda} 
+ \text{tr} \int \frac{d^4 p'}{(2 \pi)^4 i} \, \frac{1}{z_{\mathbf{p'}}}
\, f_{\mathbf{p} \mathbf{p}'} \, G_{\Lambda}^2(p') \,
\frac{d R_{\Lambda} (\mathbf{p}')}{d \Lambda} \,,
\label{eq:qpe-flow}
\end{eqnarray}
where we have used Eqs.~(\ref{eq:delta_G}),~(\ref{eq:Y-amplitude})
and~(\ref{eq:qp-int}). Combined with Eq.~(\ref{dyson-reg}), the
right-hand side of the flow equation for the quasiparticle energy,
Eq.~(\ref{eq:qpe-flow}), can be written as
\begin{equation}
\frac{d \tilde{\varepsilon}_{\mathbf{p}}}{d \Lambda} 
= z_{\mathbf{p}} \biggl( \frac{\partial 
\Sigma_{\Lambda}(\tilde{\varepsilon}_{\mathbf{p}},{\mathbf{p}})}{
\partial \Lambda} + \frac{d R_{\Lambda}(\mathbf{p})}{d \Lambda}
\biggr) \equiv \frac{d \varepsilon_{\mathbf{p}}}{d \Lambda}
+ z_{\mathbf{p}} \, \frac{d R_{\Lambda}(\mathbf{p})}{d \Lambda} \,,
\label{eq:qpe2-flow}
\end{equation}
where we have identified ${d \varepsilon_{\mathbf{p}}}/{d \Lambda}$
with $z_{\mathbf{p}} \, {\partial
\Sigma_{\Lambda}(\tilde{\varepsilon}_{\mathbf{p}},{\mathbf{p}})}/
\partial \Lambda$. In the limit $\Lambda \to 0$, $\varepsilon_{\mathbf{p}}$
approaches the solution of the Dyson equation, Eq.~(\ref{qpenergies}).

\begin{figure}[t]
\begin{center}
\includegraphics[scale=1.0]{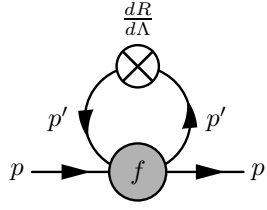}
\end{center}
\caption{Flow equation for the self-energy in the quasiparticle
approximation.\label{fig:flow-2p}}
\end{figure}

For our purpose, we use following regulator (adapted from Ref.~\cite{Morris})
\begin{equation}
R_{\Lambda} (\mathbf{p}) = \frac{\mathbf{p}^2}{2m} 
\biggl[ \frac{1}{\Theta_{\epsilon} (|\mathbf{p}| - (\kf + \Lambda)) 
+ \Theta_{\epsilon} (\kf - \Lambda - |\mathbf{p}|)} - 1 \biggr] \,,
\label{eq:reg}
\end{equation}
where $\underset{\epsilon \rightarrow 0}{\text{lim}} \, 
\Theta_{\epsilon}(x) \rightarrow \Theta(x)$ at the end of the calculation.
The regulator suppresses low-lying single-particle modes with momenta
in the range $k_F-\Lambda < |\mathbf{p}| < k_F+\Lambda$ (see
Fig.~\ref{sea}). In the limit of a sharp cutoff, $\epsilon \rightarrow
0$, one finds~\cite{Morris}
\begin{equation}
G^2_{\Lambda}(p) \, \frac{d R_{\Lambda} (\mathbf{p})}{d \Lambda}
= -\frac{\delta (|\mathbf{p}| - (\kf + \Lambda)) 
+ \delta(|\mathbf{p}| - (\kf-\Lambda))}{\omega - 
\frac{\mathbf{p}^2}{2m} - \Sigma_{\Lambda} (\omega, \mathbf{p})} \,.
\end{equation}
Keeping only the quasiparticle contribution in the second term of
Eq.~(\ref{eq:qpe-flow}) and canceling the trivial renormalization
due to the explicit regulator term in
$\tilde{\varepsilon}_{\mathbf{p}}$, we find\footnote{The resulting
flow equation, Eq.~(\ref{eq:flow-qp-energy}), is consistent with
Eq.~(\ref{eq:qp-energy-var}), if we make the natural identification
$\delta n_{\mathbf{p}} = -n_{\mathbf{p}}^0 \, \delta(|\mathbf{p}|
-(k_F-\Lambda)) \, d\Lambda$.}
\begin{equation}
\frac{d \varepsilon_{\mathbf{p}}}{d \Lambda} 
= - \text{tr} \int \frac{d^3p'}{(2 \pi)^3} \, f_{\mathbf{p} \mathbf{p}'}
\, n_{\mathbf{p}',\Lambda}^0 \Bigl[ \delta (|\mathbf{p}'| - (\kf +
\Lambda)) + \delta(|\mathbf{p}'| - (\kf-\Lambda)) \Bigr] \,.
\label{eq:flow-qp-energy}
\end{equation}
This flow equation for the quasiparticle energy,
Eq.~(\ref{eq:flow-qp-energy}), is illustrated diagrammatically in
Fig.~\ref{fig:flow-2p}.

The four-point vertex that enters the flow equation for the
self-energy is the quasiparticle-quasihole irreducible quasiparticle
interaction $f_{\mathbf{p}\mathbf{p}'}$. This result can be understood
by recognizing that the quasiparticle-quasihole reducible
contributions of the full four-point vertex $\Gamma^{\Lambda
(4)}_{p,p,p',p'}$ in Eq.~(\ref{eq:flow-one-pi}) do not contribute
for the kinematics relevant to the self-energy, for
$|\mathbf{q}|/\omega=0$. We can therefore replace the full
four-point vertex in Eq.~(\ref{eq:flow-one-pi}) by $\Gamma^\omega$,
which for quasiparticle kinematics is proportional to 
$f_{\mathbf{p}\mathbf{p}'}$.

\begin{figure}[t]
\begin{center}
\parbox{3cm}{\includegraphics[width=3.0cm]{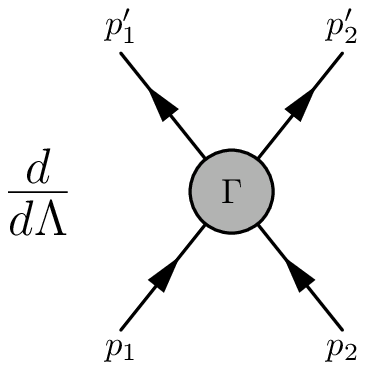}}
$\: = \:$
\parbox{3.5cm}{\includegraphics[width=3.5cm]{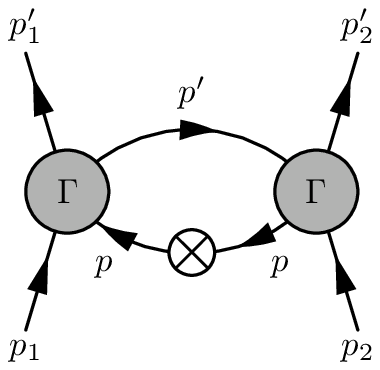}}
$\: + \:$
\parbox{3.5cm}{\includegraphics[width=3.5cm]{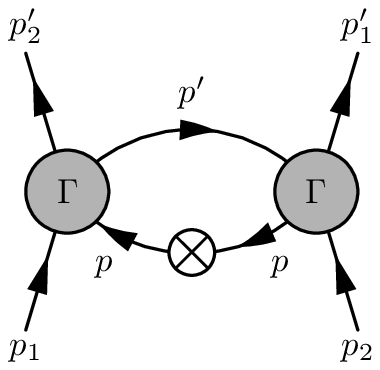}}
\\[2mm] \hspace{3.0cm} $\: + \:$ \hspace{0.1cm}
\parbox{2.5cm}{\includegraphics[height=4cm]{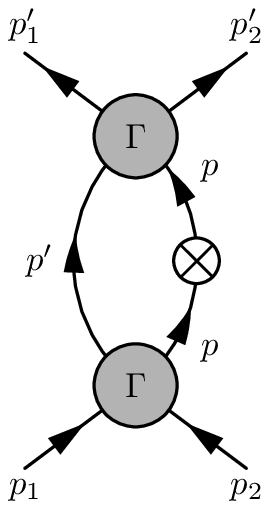}}
$\: + \:$
\parbox{3.5cm}{\includegraphics[width=2.5cm]{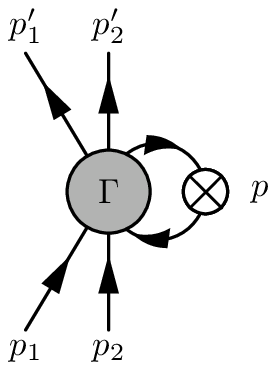}}
\end{center}
\caption{Flow equation for the four-point function $\Gamma^{(4)}$. For
each of the first three diagrams on the right-hand side, there is an 
additional diagram, where the unfilled crossed circle is inserted on 
the other internal line.\label{fig:four-point-flow}}
\end{figure}

The flow equation for the four-point function,
\begin{equation}
\Gamma^{(4)}_{p_1',p_2',p_1,p_2} = 
\frac{\delta^4 \Gamma}{\delta \psi^\ast_{p_2'} \, \delta \psi^\ast_{p_1'}
\delta \psi_{p_2} \, \delta \, \psi_{p_1}} \,,
\end{equation}
is obtained by functionally differentiating Eq.~(\ref{eq:flow}) twice
with respect to $\psi^\ast$ and twice with respect to
$\psi$.\footnote{For vanishing external sources, all vertices with an
odd number of external fermion lines vanish.} For details we refer
the reader to Ref.~\cite{Ellwanger}. The resulting flow equation is
illustrated diagrammatically in Fig.~\ref{fig:four-point-flow}.  There
are two types of contributions to the flow equation: those involving
two four-point functions (where all three channels, the particle-hole
ZS and ZS$'$ channels, as well as the particle-particle/hole-hole BCS
channel contribute) and one obtained by closing two legs of
the six-point function.

The particle-hole channels take into account contributions to the
quasiparticle scattering amplitude and interaction due to long-range
density and spin-density excitations, whereas the BCS channel builds
up contributions from the coupling to high-momentum states and due to
pairing correlations. For the application to neutron matter, we start
the many-body calculation from low-momentum interactions
$\vlowk$~\cite{Vlowk}, for which the particle-particle channel is
perturbative at nuclear densities, except for low-lying pairing
correlations~\cite{matter,Weinberg}. This is in contrast to hard
potentials, where the coupling to high momenta renders all channels
non-perturbative.

We therefore solve the flow equations for the four-point function
shown in Fig.~\ref{fig:four-point-flow} including only the
particle-hole contributions (the first and second diagrams). After the
RG flow, we calculate the low-lying pairing correlations by solving
the quasiparticle BCS gap equation. In our first study~\cite{RGnm}, we
neglected the contribution from the six-point function to the flow
equation (the last term in Fig.~\ref{fig:four-point-flow}) and
approximated the internal Green's functions by the quasiparticle part
(the first term on the right-hand-side of Eq.~(\ref{G_full})). The
resulting flow equations for the quasiparticle scattering amplitude
\begin{equation}
a(\mathbf{q},\mathbf{q}'; \Lambda) = z_{\kf}^2 \,
\Gamma^{(4)}_{p-\frac{q}{2},p'+\frac{q}{2},p+\frac{q}{2},p'-\frac{q}{2}}
\Bigl|_{\omega = \omega' = \varepsilon_{\rm F}, \, q_0 =0} \,,
\end{equation}
and the quasiparticle interaction $f(\mathbf{q},\mathbf{q}';\Lambda)$
are given by~\cite{RGnm}:
\begin{eqnarray}
\frac{d}{d\Lambda} a(\mathbf{q},\mathbf{q}';\Lambda) 
&=& z_{\kf}^2 \frac{d}{d\Lambda} 
\left[ g \int_{\text{fast},\Lambda} \frac{d^3 \mathbf{p}''}{(2 \pi)^3} 
\frac{n_{\mathbf{p}''+\mathbf{q}/2} - n_{\mathbf{p}''-\mathbf{q}/2}}{
\varepsilon_{\mathbf{p}''+\mathbf{q}/2} - 
\varepsilon_{\mathbf{p}''-\mathbf{q}/2}} \right] \label{eq:RG_longrange1} \\
&\times& a \Big( \mathbf{q}, \frac{\mathbf{p}+\mathbf{p}'}{2} + 
\frac{\mathbf{q}'}{2} - \mathbf{p}''; \Lambda \Big) 
a\big( \mathbf{q}, \mathbf{p}'' - \frac{\mathbf{p}+\mathbf{p}'}{2}
+\frac{\mathbf{q}'}{2}; \Lambda \big) \nonumber \\
&+& \frac{d}{d \Lambda} f(\mathbf{q},\mathbf{q}';\Lambda) \,,
\nonumber \\[1mm]
\frac{d}{d \Lambda} f(\mathbf{q},\mathbf{q}';\Lambda) 
&=& - z_{\kf}^2 \frac{d}{d \Lambda} \left[ g \int_{\text{fast},\Lambda}
\frac{d^3 \mathbf{p}''}{(2 \pi)^3} \frac{n_{\mathbf{p}''+\mathbf{q}'/2} -
n_{\mathbf{p}''-\mathbf{q}'/2}}{\varepsilon_{\mathbf{p}''+\mathbf{q}'/2} 
- \varepsilon_{\mathbf{p}''-\mathbf{q}'/2}}
\right] \label{eq:RG_longrange2} \\
&\times& a \big( \mathbf{q}', \frac{\mathbf{p}+\mathbf{p}'}{2} 
+ \frac{\mathbf{q}}{2} - \mathbf{p}''; \Lambda \big) 
a\big( \mathbf{q}', \mathbf{p}'' - \frac{\mathbf{p}+\mathbf{p}'}{2}
+\frac{\mathbf{q}}{2}; \Lambda \big)\,. \nonumber
\end{eqnarray}
Here the spin labels and the spin trace in the flow equation have been
suppressed. In a spin-saturated system the spin dependence of $f$ and
$a$ is of the form of Eq.~(\ref{eq:spin}), when non-central forces are
neglected. We note that for $\mathbf{q} = \mathbf{q}' = 0$ and
identical spins the contributions from the ZS and ZS$'$ channels to
the scattering amplitude $a$ cancel, as required by the Pauli
principle. Thus, the resulting quasiparticle interaction and the
corresponding Fermi liquid parameters satisfy the Pauli-principle sum
rules, see Eq.~(\ref{sumrule}). Moreover, the flow equations,
Eqs.~(\ref{eq:RG_longrange1}) and~(\ref{eq:RG_longrange2}), yield the
correct quasiparticle-quasihole reducibility of the scattering
amplitude $a(\mathbf{q},\mathbf{q}'; \Lambda)$ and of the
quasiparticle interaction $f(\mathbf{q},\mathbf{q}'; \Lambda)$.

At an initial scale $\Lambda = \Lambda_0$, the quasiparticle
scattering amplitude and interaction start from the free-space
interaction. In neutron matter, non-central and three-nucleon forces
are weaker, and we therefore solved the flow equations starting from
low-momentum two-nucleon (NN) interactions $\vlowk$~\cite{Vlowk},
\begin{equation}
a(\mathbf{q}, \mathbf{q'}; \Lambda_0) = f(\mathbf{q}, \mathbf{q'};
\Lambda_0) = V_{\text{low}\,k} (\mathbf{q}, \mathbf{q}') \,,
\label{eq:initial_a}
\end{equation}
including only scalar and spin-spin interactions, which dominate
at densities below nuclear saturation density.

\begin{figure}[t]
\begin{center}
\raisebox{0.45cm}{\includegraphics[height=1.0cm]{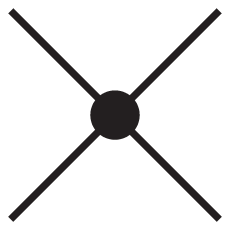}}
\hspace{1.0cm}
\includegraphics[height=2.0cm]{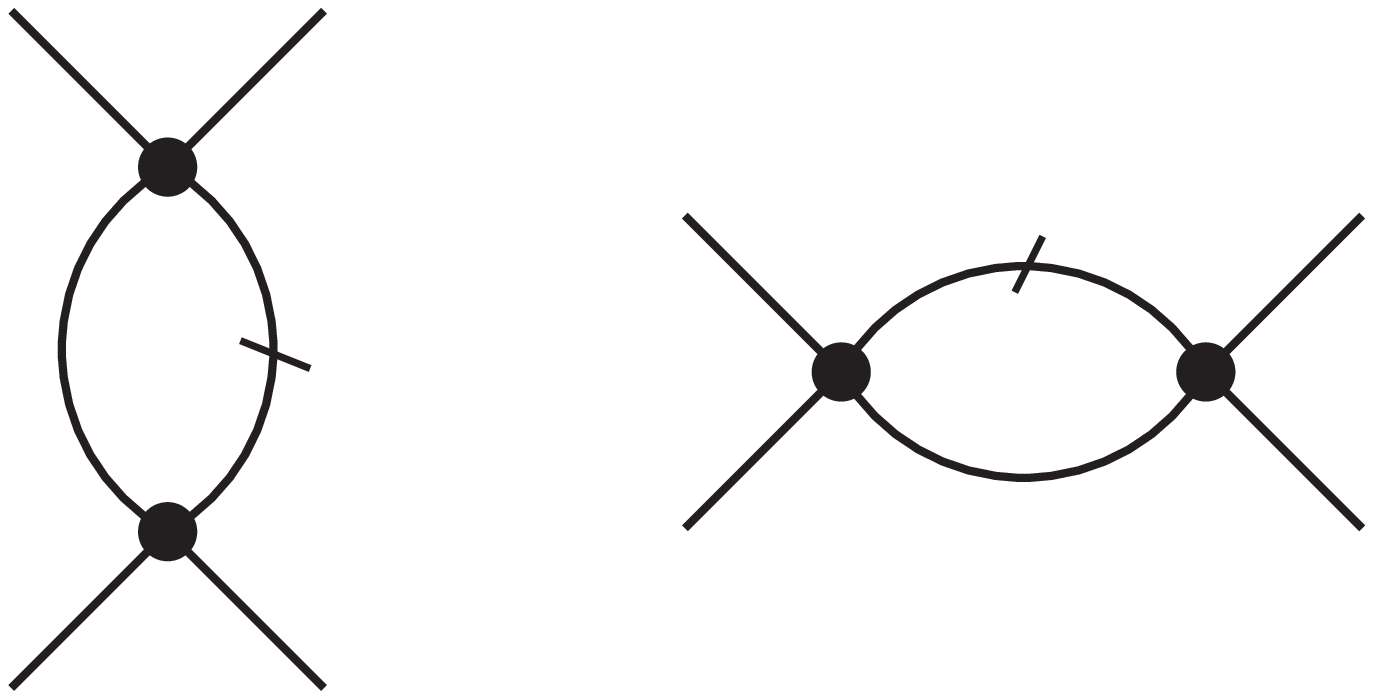}
\end{center}
\caption{The free-space two-body interaction, shown by the left
diagram, is the lowest order contribution to the quasiparticle
scattering amplitude. The two one-loop particle-hole diagrams on the
right are generated by the RG flow in the first iteration. There are
two more diagrams, where the slash is on the other internal
line.\label{fig:parquet0}}
\end{figure}
 
It is instructive to study the RG method diagrammatically, to
understand how many-body correlations are generated by the flow
equations. We discuss the set of diagrams, which is generated by the
flow equation for the scattering amplitude,
Eq.~(\ref{eq:RG_longrange1}). We denote the antisymmetrized free-space
two-body interaction $\Gamma_{\text{vac}}$, the initial condition for
the flow equation, by a dot (see Fig.~\ref{fig:parquet0}). Starting
from $\Gamma_{\text{vac}}$, we integrate out the first shell $\delta
\Lambda$ of high-lying particle-hole excitations to obtain the
effective scattering amplitude at the lower scale $\Lambda_1 =
\Lambda_0 - \delta\Lambda$. The RG flow includes contributions from
both particle-hole channels, which leads to the four diagrams, two of
which are shown in Fig.~\ref{fig:parquet0}. The lines marked by a
slash are restricted by the regulator to momenta in the shell
$\Lambda_1 \leqslant k < \Lambda_0$.

\begin{figure}[t]
\begin{center}
\includegraphics[height=2.0cm]{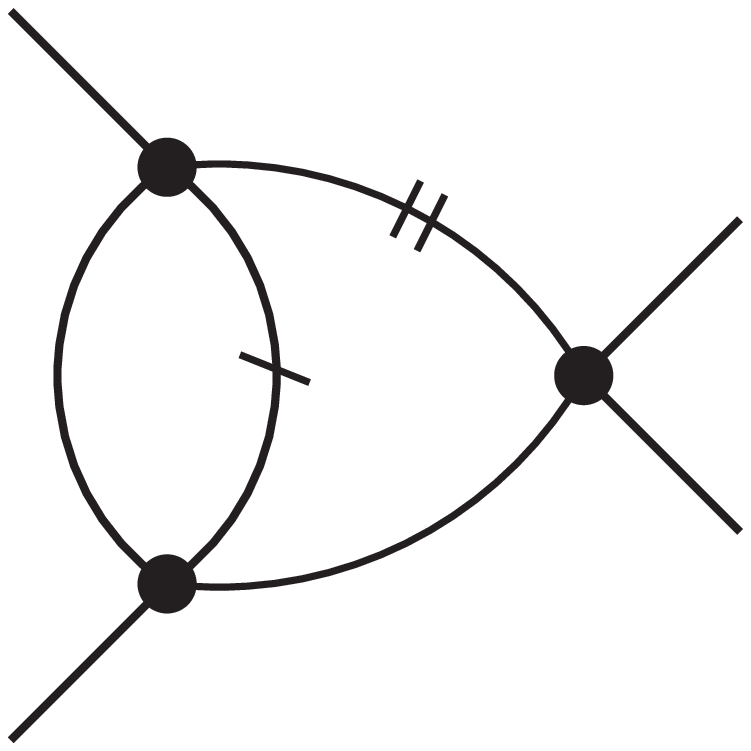}
\quad
\includegraphics[height=2.0cm,origin=c,angle=-180]{third-order1.eps}
\quad
\includegraphics[height=2.0cm]{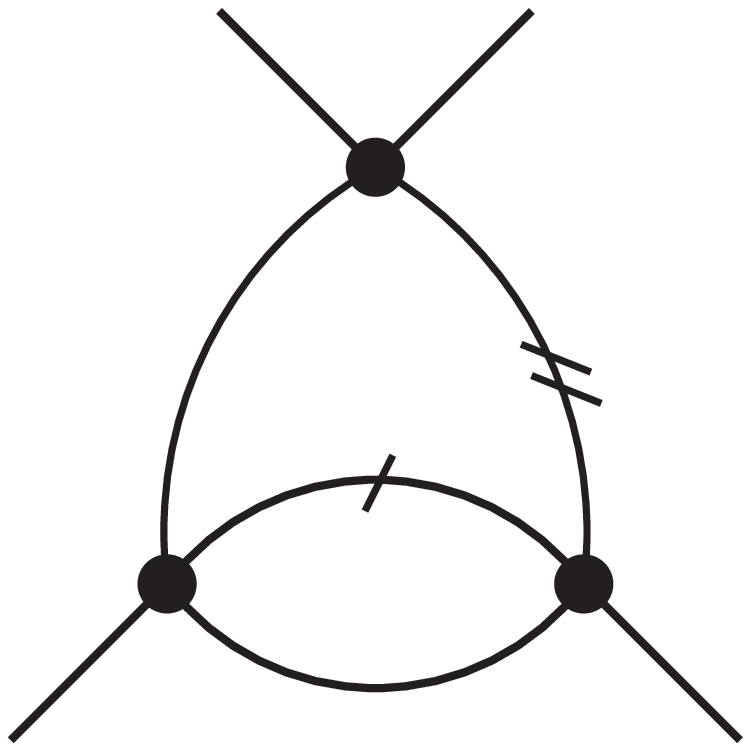}
\quad
\includegraphics[height=2.0cm,origin=c,angle=-180]{third-order2.eps}
\\[2mm]
\includegraphics[height=3.0cm]{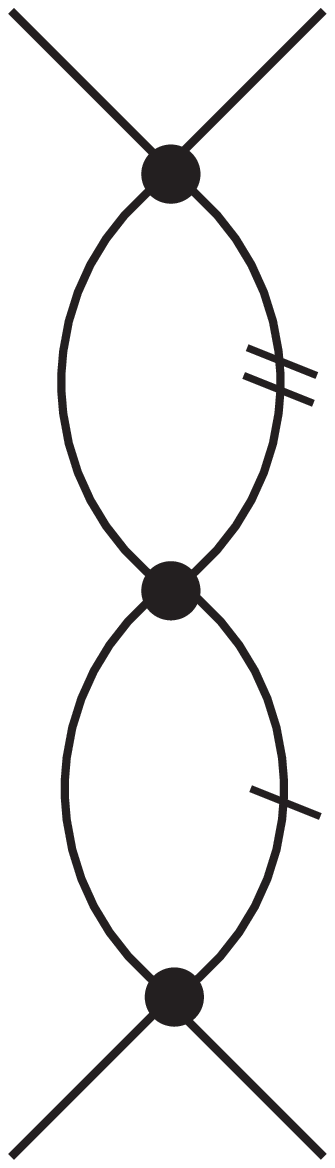}
\qquad
\includegraphics[height=3.0cm,origin=c,angle=-180]{third-order3.eps}
\quad
\raisebox{1.0cm}{\includegraphics[height=1.0cm]{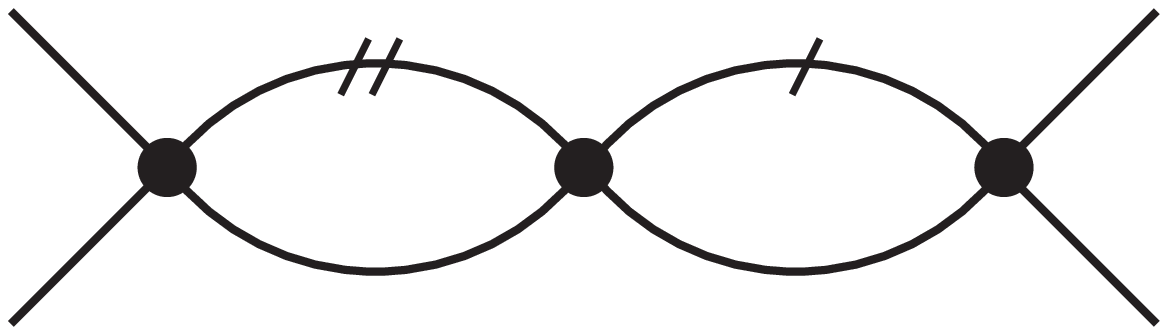}
\quad
\includegraphics[height=1.0cm,origin=c,angle=-180]{third-order4.eps}}
\end{center}
\caption{Two-loop diagrams that contribute to the scattering amplitude 
after two iterations of the flow equations.\label{fig:parquet2}}
\end{figure}

The four-point vertex used in each iteration of the flow equation is
the one obtained in the previous iteration. Thus, in the second
iteration, one finds the one-loop diagrams shown in
Fig.~\ref{fig:parquet0}, but now with the momentum of the marked line
in the second shell $\Lambda_2 = \Lambda_1 - \delta \Lambda \leqslant
k < \Lambda_1$. In addition, the RG flow generates the two-loop
diagrams shown in Fig.~\ref{fig:parquet2}. Here, lines with momenta in
the first shell are marked by one slash and those with momenta in the
second shell by two slashes. For every diagram shown, there are three
more diagrams obtained by moving the slash or the two slashes from a
particle (hole) line to the corresponding hole (particle) line. The
first four diagrams illustrate the coupling between the two
particle-hole channels generated by the RG equations.

\begin{figure}[t]
\begin{center}
\includegraphics[height=2.0cm]{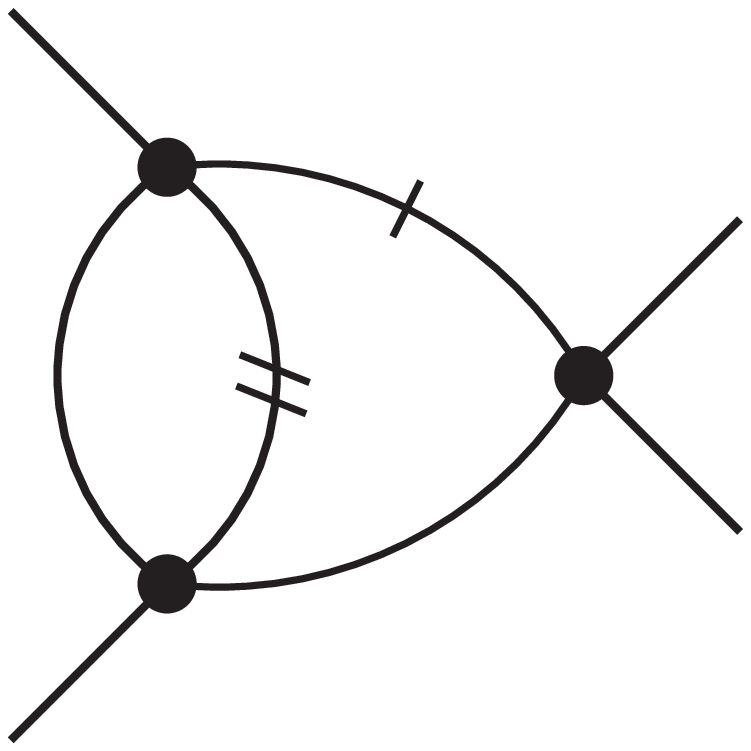}
\quad
\includegraphics[height=2.0cm,origin=c,angle=180]{third-order5.eps}
\quad
\includegraphics[height=2.0cm]{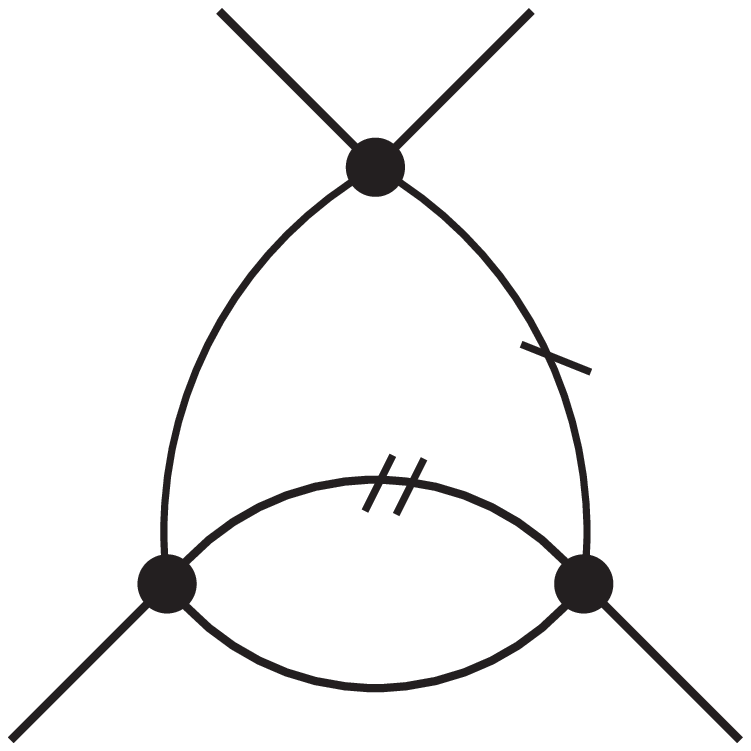}
\quad
\includegraphics[height=2.0cm,origin=c,angle=180]{third-order6.eps}
\end{center}
\caption{The two-loop planar diagrams that are not generated by the
RG flow after two iterations.\label{fig:parquet3}}
\end{figure}

We observe that the four diagrams of third order in the vacuum
interaction shown in Fig.~\ref{fig:parquet3} are missing in this
scheme. These diagrams are obtained when the last term in
Fig.~\ref{fig:four-point-flow} (the contribution of the six-point
function to the flow of the four-point function) is included. This
shows that the flow equations reproduce the full one-loop
particle-hole phase space exactly, augmented by a large antisymmetric
subset of the particle-hole parquet diagrams. In a truncation scheme
where the BCS channel and the six-point function in
Fig.~\ref{fig:four-point-flow} is included, the RG sums all
planar diagrams using the RG.

\subsection{Fermi liquid parameters and scattering amplitude}

\begin{figure}[t]
\begin{center}
\includegraphics[scale=0.26,clip=]{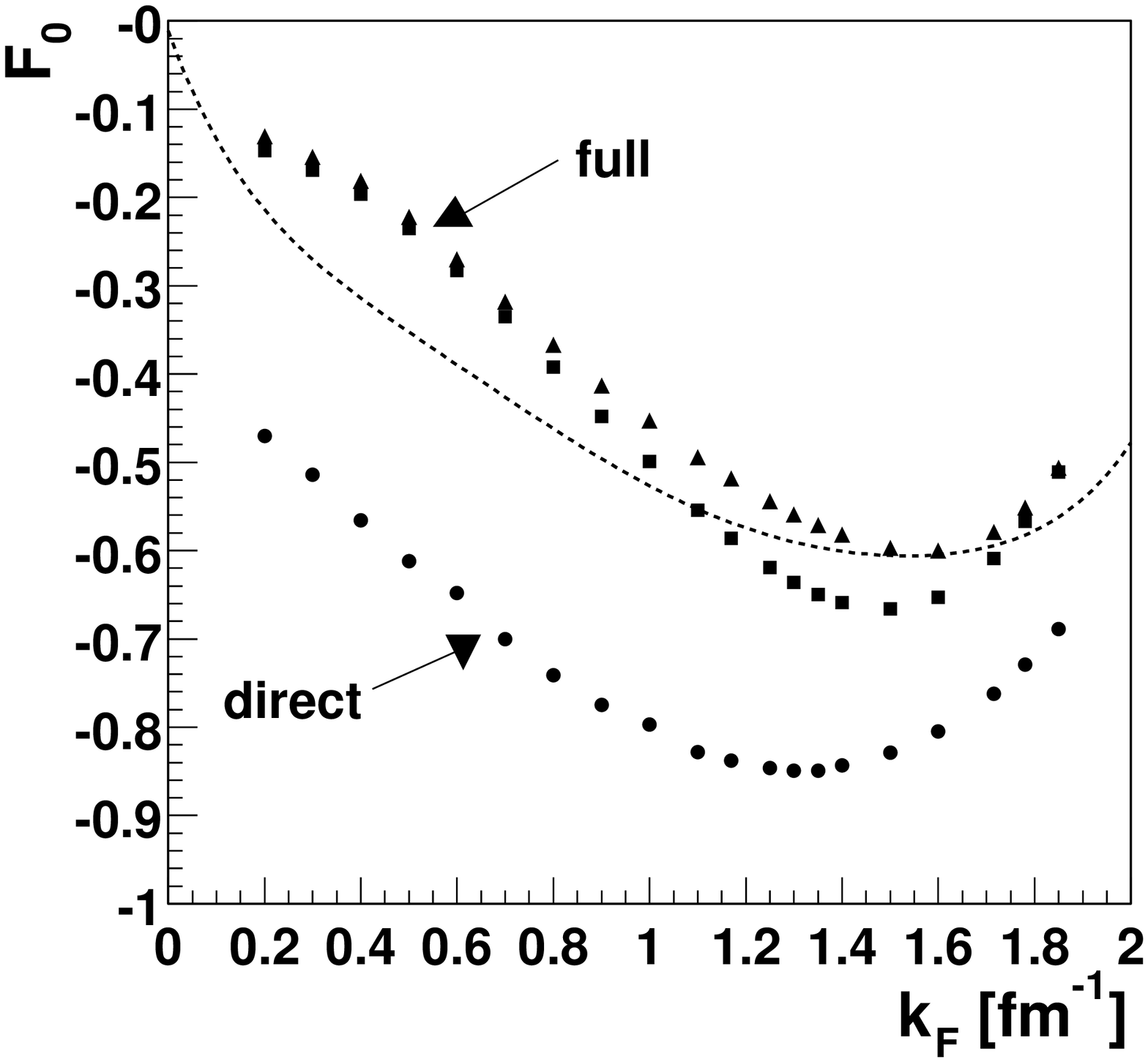}
\includegraphics[scale=0.26,clip=]{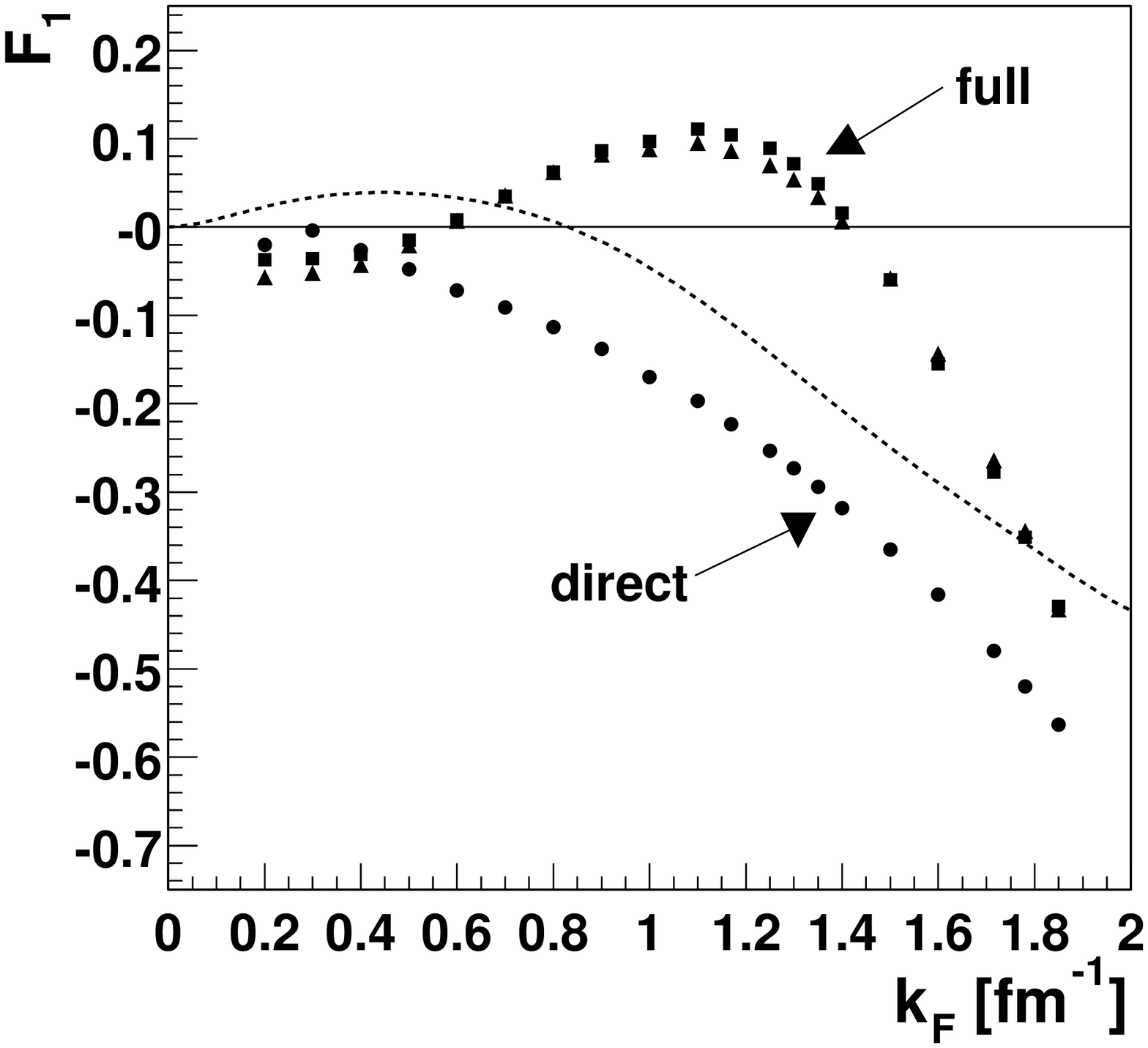}
\includegraphics[scale=0.26,clip=]{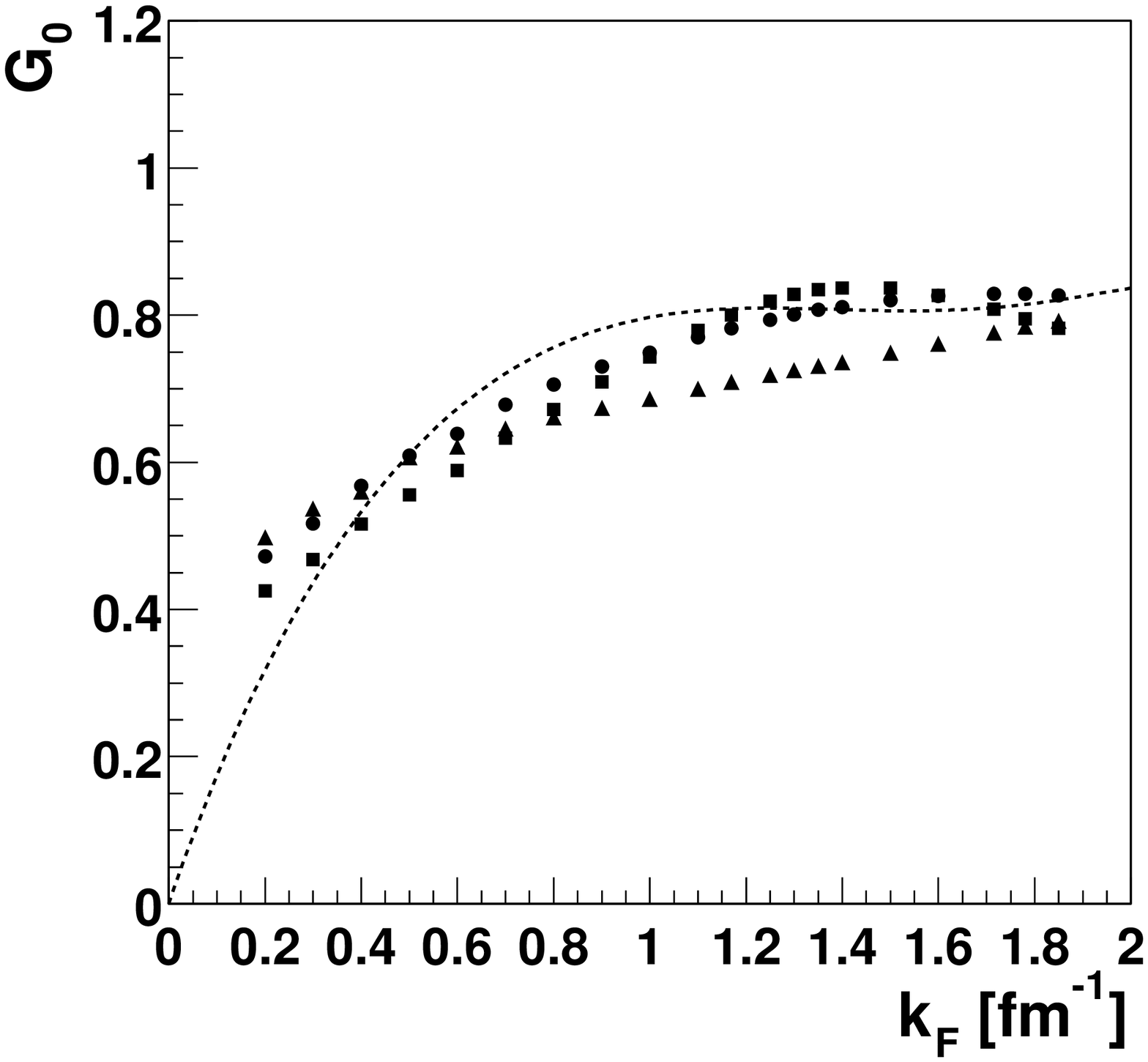}
\includegraphics[scale=0.26,clip=]{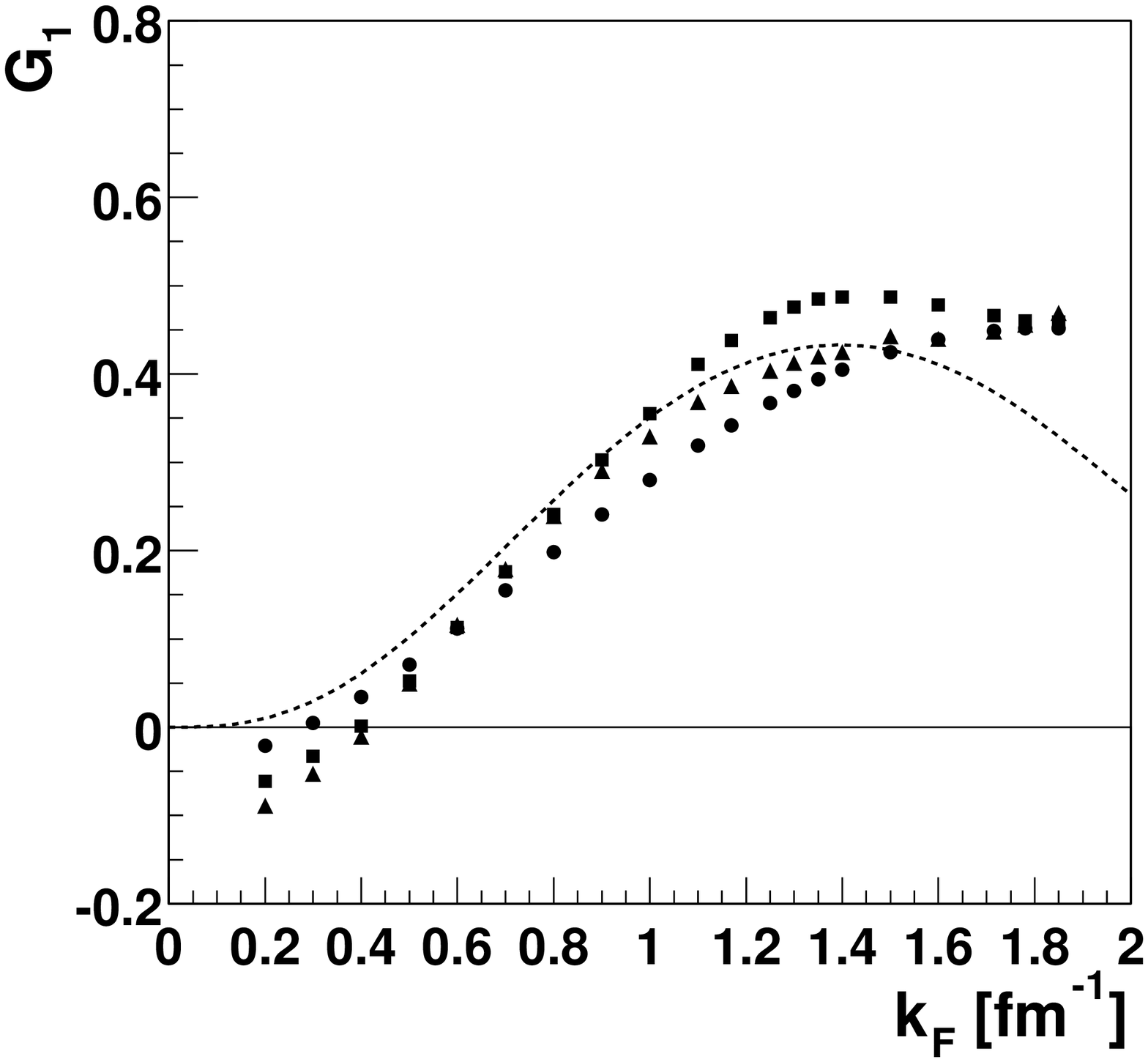}
\end{center}
\caption{The $l=0$ and $l=1$ Fermi liquid parameters versus the Fermi
momentum $\kf$. The dots denote the direct contribution only (\zf=1,
but including the effective mass in the density of states), whereas
the squares (static \zf~factor) and the triangles (adaptive
\zf~factor) are calculated from the full RG solution. The results of
Wambach {\em et al.}~\cite{WAP2} are given for comparison as
dashed lines.}
\label{Flp}
\end{figure}

In Fig.~\ref{Flp}, we show the resulting $l=0$ and $l=1$ Fermi liquid
parameters as a function of the Fermi momentum $\kf$. In this section
we use the notation common in nuclear physics,
\begin{eqnarray}
f_{\mathbf{p} \sigma \mathbf{p'} \sigma'} &=& f_{\mathbf{p} \mathbf{p}'}
+ g_{\mathbf{p} \mathbf{p}'} \, \boldsymbol{\sigma} \cdot
\boldsymbol{\sigma}' \,, \\[1mm]
a_{\mathbf{p} \sigma \mathbf{p'} \sigma'} &=& a_{\mathbf{p} \mathbf{p}'}
+ b_{\mathbf{p} \mathbf{p}'} \,\boldsymbol{\sigma} \cdot
\boldsymbol{\sigma}' \,,
\end{eqnarray}
with $F_l=N_0 \, f_l$ and $G_l=N_0 \, g_l$. The cutoff scale of the
starting $\vlowk$ is taken as $\Lambda=\sqrt{2}\,\kf$. This choice has
the advantage that scattering to high-lying states in the
particle-particle channel is uniformly accounted for at different
densities. The low-momentum interaction $\vlowk$ then drives the flow
in the ZS$'$ channel for the quasiparticle interaction, and thus our
results include effects of the induced interaction. A further
advantage of the RG approach is that the calculations can be performed
without truncating the expansion of the quasiparticle interaction,
Eq.~(\ref{eq:f_expansion}), at some $l$.

The flow equations were solved with two different assumptions for the
\zf~factor. In one case, we use a static, density-independent mean
value of $z_{k_F}^2 = 0.9$ which remains unchanged under the RG. In
the other case, we compute the \zf~factor dynamically, by assuming
that the change of the effective mass from the initial one, based on
the momentum dependence of $\vlowk$, is due to the \zf~factor
alone~\cite{RGnm}. Generally, we find a very good agreement between
our results and the ones obtained using the polarization potential
model by Ainsworth {\em et al.}~\cite{WAP1,WAP2}. There are minor
differences in the value of the effective mass, which is treated
self-consistently in the RG approach. We note that, in the density
range $0.6 \, \text{fm}^{-1} < \kf < 1.4 \, \text{fm}^{-1}$, we find
that the effective mass at the Fermi surface exceeds unity. The
quasiparticle interaction was also calculated taking into account
induced interactions~\cite{FLnm,JKMS,Lombardo}. In these papers, the
value for the $l=0$ spin-dependent parameter $G_0$ is in good
agreement with ours, while there are differences for the
spin-independent $F_0$. We stress the important role of the large
$G_0$ for the induced interaction. This Landau parameter causes the
strong spin-density correlations, which in turn enhance the Landau
parameter $F_0$ and consequently the incompressibility of neutron
matter.

The results can be qualitatively understood by inspecting the explicit
spin dependence of the RG equation for the quasiparticle interaction,
Eqs.~(\ref{eq:RG_longrange1}) and (\ref{eq:RG_longrange2}) for $q=0$,
\begin{align}
\Lambda\,\frac{d}{d \Lambda} \, a(q=0,\qp;\Lambda) &= 
- \Theta(q'-2\Lambda) \: \big(
\frac{1}{2} \, \beta_{\text{ZS}'}[a,\qp,\Lambda] + \frac{3}{2} \,
\beta_{\text{ZS}'}[b,\qp,\Lambda] \big) \,, \label{qpfrg} \\[2mm]
\Lambda\,\frac{d}{d \Lambda} \, b(q=0,\qp;\Lambda) &= 
- \Theta(q'-2\Lambda) \: \big(
\frac{1}{2} \, \beta_{\text{ZS}'}[a,\qp,\Lambda] - \frac{1}{2} \,
\beta_{\text{ZS}'}[b,\qp,\Lambda] \big)\,, \label{qpgrg}
\end{align}
where we have introduced the $\beta$ functions
$\beta_{\text{ZS}'}[a,\qp;\Lambda]$ and
$\beta_{\text{ZS}'}[b,\qp;\Lambda]$ for the contribution from density
and spin-density fluctuations in the ZS$'$ channel to the RG flow. In
this qualitative argument we neglect Fermi liquid parameters with $l
\geqslant 1$. The flow equations, Eqs.~(\ref{eq:RG_longrange1}) and
(\ref{eq:RG_longrange2}), show that the $\beta$ functions are quadratic
in the four-point functions $a$ and $b$, while Eq.~(\ref{eq:initial_a})
implies that the initial values for $a$ and $b$ are given by the
lowest order contribution to the quasiparticle interaction.  At a
typical Fermi momentum $\kf = 1.0\,\text{fm}^{-1}$, we observe that
the initial $F_0$ and $G_0$ are similar in absolute value, $|F_0|
\approx |G_0| \approx 0.8$. Consequently, there is a cancellation
between the contributions due to the spin-independent and
spin-dependent parameters in Eq.~(\ref{qpgrg}), while in
Eq.~(\ref{qpfrg}) both contributions are repulsive. Thus, one expects
a relatively small effect of the RG flow on $G_0$ and a substantial
renormalization of $F_0$, in agreement with our results.

The RG approach enables us to compute the scattering amplitude for
general scattering processes on the Fermi surface, without making
further assumptions for the dependence on the particle-hole momentum
transfers $q$ and $q'$. The scattering amplitude at finite momentum
transfer is of great interest for calculating transport processes and
superfluidity, as discussed in the following section.  The flow
equations treat the dependence on the momenta $q$ and $q'$ on an equal
footing and maintain the symmetries of the scattering amplitude. For
scattering on the Fermi surface, ${\bf q}$, ${\bf q'}$ and ${\bf P}$
are orthogonal, and they are restricted to $q^2+q^{\prime 2}+P^2
\leqslant 4 \, \kf^2$. Therefore, in Ref.~\cite{RGnm} we approximated
$a({\bf q},{\bf q'};\Lambda) = a(q^2,q'^2;\Lambda)$ for the solution of
the flow equations to extrapolate off the Fermi surface. On the
$\vlowk$ level, we checked that the ${\bf q} \cdot {\bf q'}$
dependence is small for neutrons.

\subsection{Superfluidity in neutron matter}

Superfluidity plays a key role in strongly-interacting many-body
systems. Pairing in infinite matter impacts the cooling of isolated
neutron stars~\cite{slow} and of the neutron star crust~\cite{crust},
and is used to develop non-empirical energy-density
functionals~\cite{Lesinski}. In this section, we discuss superfluidity
in neutron matter, with particular attention to induced interactions
using the RG approach.

\begin{figure}[t]
\begin{center}
\includegraphics[clip=,width=4.0in]{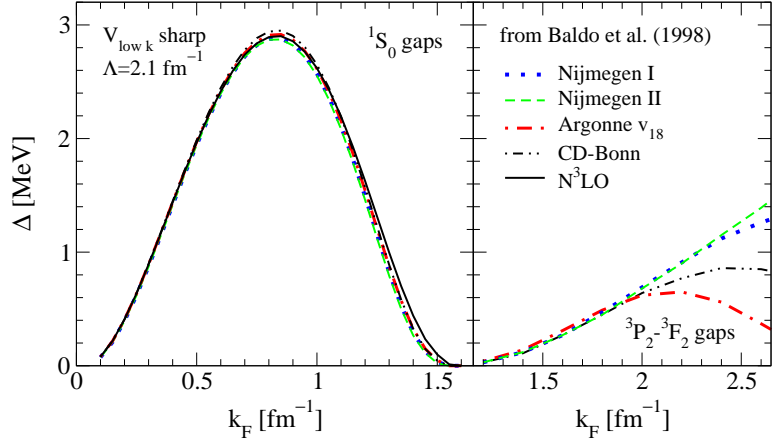}
\end{center}
\caption{The $^1$S$_0$ (left panel) and $^3$P$_2-^3$F$_2$ (right
panel) superfluid pairing gaps $\Delta$ at the Fermi surface as a
function of Fermi momentum $\kf$ in neutron matter. The gaps are
obtained from charge-dependent NN interactions at the BCS level.
For details see Refs.~\cite{Kai,Baldo}.\label{BCSgaps}}
\end{figure}

Figure~\ref{BCSgaps} shows the superfluid pairing gaps in neutron
matter, obtained by solving the BCS gap equation with a free spectrum.
In this approximation, the $^1$S$_0$ superfluid gap $\Delta(k)$ is
determined from the gap equation,
\begin{equation}
\Delta(k) = - \frac{1}{\pi} \int dp \, p^2 \: \frac{V(k,p) \,
\Delta(p)}{\sqrt{\xi^2(p) + \Delta^2(p)}} \,,
\label{gapeq}
\end{equation}
where $V(k,p)$ is the free-space NN interaction in
the $^1$S$_0$ channel, $\xi(p) \equiv p^2/(2m) - \mu$, and for a free
spectrum the chemical potential is given by $\mu = \kf^2/(2m)$. At low
densities (in the crust of neutron stars), neutrons form a $^1$S$_0$
superfluid. At higher densities, the S-wave interaction is repulsive
and neutrons pair in the $^3$P$_2$ channel (with a small coupling to
$^3$F$_2$ due to the tensor force). Figure~\ref{BCSgaps} demonstrates
that the $^1$S$_0$ BCS gap is practically independent of nuclear
interactions, and therefore strongly constrained by NN phase
shifts~\cite{Kai}. This includes a very weak cutoff dependence for
low-momentum interactions $\vlowk$ with sharp or sufficiently narrow
smooth regulators with $\lm > 1.6 \fmi$. The inclusion of N$^2$LO
three-nucleon (3N) forces leads to a reduction of the $^1$S$_0$ BCS
gap for Fermi momenta $\kf > 0.6 \fmi$, where the gap is
decreasing~\cite{matter}. Two-nucleon interactions are well
constrained by scattering data for relative momenta $k \lesssim 2
\fmi$~\cite{Vlowk}. The model dependencies at higher momenta show up
prominently in Fig.~\ref{BCSgaps} in the $^3$P$_2-^3$F$_2$ gaps for
Fermi momenta $\kf > 2 \fmi$~\cite{Baldo}.

\begin{figure}[t]
\begin{center}
\includegraphics[clip=,width=3.4in]{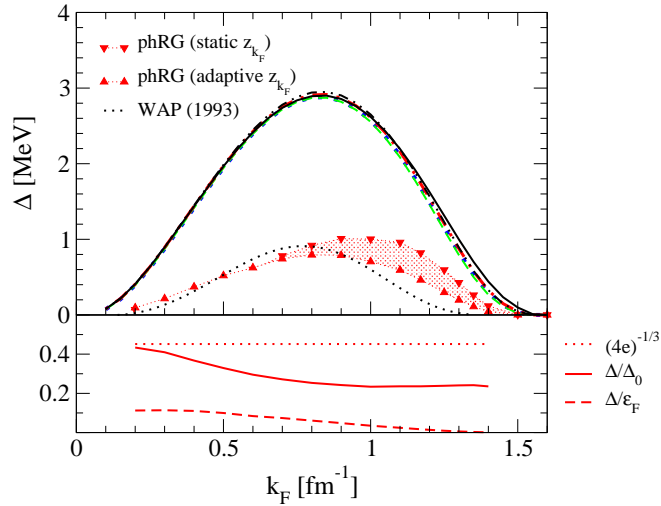}
\end{center}
\caption{Top panel: Comparison of the $^1$S$_0$ BCS gap to the
results including polarization effects through the particle-hole RG
(phRG), for details see Ref.~\cite{RGnm}, and to the results of Wambach 
{\it et al.}~\cite{WAP2}. Lower panel: Comparison of the full 
superfluid gap $\Delta$ to the BCS gap $\Delta_0$ and to the Fermi
energy $\varepsilon_{\rm F}$.\label{fullgaps}}
\end{figure}

Understanding many-body effects beyond the BCS level constitutes an
important open problem. For recent progress and a survey of results,
see for instance Ref.~\cite{Gezerlis}. At low densities, induced
interactions due to particle-hole screening and vertex corrections are
significant even in the perturbative $\kf a$ limit~\cite{Gorkov} and
lead to a reduction of the S-wave gap by a factor $(4e)^{-1/3} \approx
0.45$,
\bea
\frac{\Delta}{\varepsilon_{\rm F}} &=& \frac{8}{e^2} \, \exp \biggr\{ \biggr(
\, \begin{minipage}{4.4cm}
\includegraphics[scale=0.5,clip=]{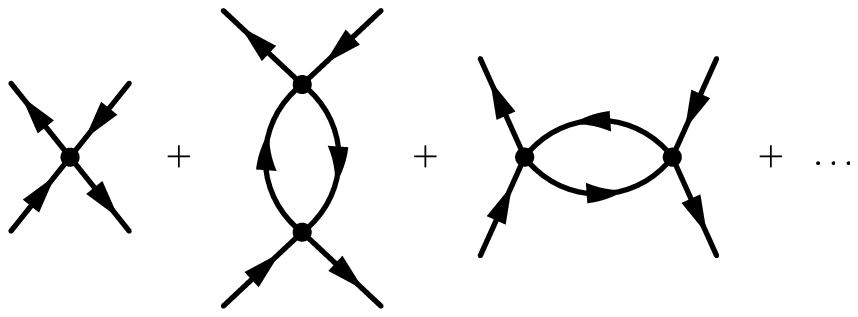}
\end{minipage}\biggr)^{-1} \biggr\} \nonumber \\
&=& (4e)^{-1/3} \, \frac{8}{e^2} \, \exp \biggl\{ \frac{\pi}{2 \kf a} + 
{\mathcal O}(\kf a)\biggr\} \,.
\label{Gorkov}
\eea
This reduction is due to spin fluctuations, which are repulsive for
spin-singlet pairing and overwhelm attractive density
fluctuations.\footnote{In finite systems, the spin and density
response differs. In nuclei with cores, the low-lying response is
due to surface vibrations. Consequently, induced interactions may
be attractive, because the spin response is weaker~\cite{Milan}.}

Here, we discuss the particle-hole RG (phRG) approach to this problem
using the BCS-channel-irreducible quasiparticle scattering amplitude
as pairing interaction~\cite{RGnm}. The quasiparticle scattering
amplitude is obtained, as discussed in the previous section, by
solving the flow equations in the particle-hole channels,
Eqs.~(\ref{eq:RG_longrange1}) and~(\ref{eq:RG_longrange2}), starting
from $\vlowk$. This builds up many-body correlations from successive
momentum shells, on top of an effective interaction with particle/hole
polarization effects from all previous shells, and thereby efficiently
includes induced interactions to low-lying states in the vicinity of
the Fermi surface beyond a perturbative calculation.

The results for the $^1$S$_0$ gap are shown in Fig.~\ref{fullgaps},
where induced interactions lead to a factor 3--4 reduction to a
maximal gap $\Delta \approx 0.8 \mev$~\cite{RGnm}.  Similar values to
those of Wambach {\it et al.}~\cite{WAP2} are found. In addition, for
the lower densities, the phRG is consistent with the dilute
result\footnote{For $\kf \approx 0. 4 \fmi$, neutron matter is close
to the universal regime, but theoretically simpler due to an
appreciable effective range $\kf r_{\rm e} \approx 1$~\cite{dEFT}.}
$\Delta/\Delta_0 = (4e)^{-1/3}$, and at the larger densities the
dotted band indicates the uncertainty due to an approximate
self-energy treatment.

Non-central spin-orbit and tensor interactions are crucial for
$^3$P$_2-^3$F$_2$ superfluidity. Without a spin-orbit interaction,
neutrons would form a $^3$P$_0$ superfluid instead.  The first
perturbative calculation of non-central induced interactions shows
that $^3$P$_2$ gaps below $10 \, {\rm keV}$ are possible (while
second-order contributions to the pairing interaction are not
substantial $| V_{\rm ind} / \vlowk | < 0.5$)~\cite{tensor}. This
arises from a repulsive induced spin-orbit interaction due to the
mixing with the stronger spin-spin interaction. As a result, neutron
P-wave superfluidity (in the interior of neutron stars) may be reduced
considerably below earlier estimates. This implies that low-mass
neutron stars cool slowly~\cite{slow}.  Smaller values for the
$^3$P$_2$ gap compared to Fig.~\ref{BCSgaps} are also required for
consistency with observations in a minimal cooling
scenario~\cite{CasA}.

\section{Outlook}

In these lecture notes we discussed RG approaches to Fermi liquids,
based on the ideas of Shankar~\cite{Shankar} and
Polchinski~\cite{Polchinski}, and formulated the flow equation in the
framework of a functional RG~\cite{Wetterich93}. The appeal of RG
methods is that they allow a systematic study of non-perturbative
correlations at different length scales. These lecture notes show that
the RG approach developed for neutron matter in Ref.~\cite{RGnm} is
equivalent to a functional RG. We reviewed results for the
quasiparticle interaction and the superfluid pairing gaps in neutron
matter, starting from low-momentum interactions and focusing on the
effects of induced interactions due to long-range particle-hole
fluctuations.

There are many important problems for RG methods applied to
many-nucleon systems. These include the role of non-central
interactions and of many-body forces. For example, we recently showed
that novel non-central interactions are generated by polarization of
the nuclear medium~\cite{tensor}. A non-perturbative RG treatment of
these effects, which have important consequences for neutron stars, is
of great interest. In recent years an alternative RG scheme, the
similarity RG (SRG)~\cite{SRG,SRGnuc} has emerged as a powerful tool
for exploring the scale dependence of nuclear interactions and the
role of many-body forces in nuclear systems. Moreover, the SRG can be
applied directly to solve the many-body problem~\cite{IMSRG}. The SRG
approach is formulated as a continuous matrix transformation acting on
the Hamiltonian. An important open question is to understand the
relation of the SRG to functional and field theory methods.

\section*{Acknowledgments}

This work was supported in part by NSERC and by the Alliance Program
of the Helmholtz Association (HA216/EMMI).

\end{document}